# Cr resonant impurity for studies of band inversion and band offsets in IV-VI semiconductors

A. Królicka [1]*, K. Gas [1,2,3], W. Dobrowolski [1], H. Przybylińska[1], Y. K. Edathumkandy[1],

J. Korczak [1,4], E. Łusakowska [1], R. Minikayev [1], A. Reszka[1],

R. Jakieła [1], L. Kowalczyk[1], A. Mirowska[5], M. Gryglas-Borysiewicz[6], J. Kossut[1],

M. Sawicki [1,3], A. Łusakowski [1], P. Bogusławski [1], T. Story[1,4], K. Dybko [1,4]

[1] *Institute of Physics, Polish Academy of Sciences, al. Lotników 32/46, 02668 Warsaw, Poland*
[2] *Center for Science and Innovation in Spintronics, Tohoku University, 2-1-1Katahira, Aoba-ku, Sendai 980-8577, Japan*
[3] *Research Institute of Electrical Communication, Tohoku University, 2-1-1Katahira Aoba-ku, Sendai 980-8577, Japan*
[4] *International Research Centre MagTop, Institute of Physics Polish Academy of Sciences, al. Lotników 32/46, 02668 Warsaw, Poland*
[5] *ENSEMBLE3 Centre of Excellence for Nanophotonics, Advanced Materials and Novel Crystal growth-based Technologies, Wólczyńska 133, 01919 Warsaw, Poland*
[6] *Faculty of Physics, University of Warsaw, Pasteura 5, 02093 Warsaw, Poland*

\*   Corresponding author: krolicka@ifpan.edu.pl

**Abstract:**

Understanding the electronic structure of transition-metal dopants in IV–VI semiconductors is critical for tuning their band structure. We analyze properties of the Cr dopant in $Pb_{1-x}Sn_xTe$ and PbSe by magnetic and transport measurements, which are interpreted based on density functional theory calculations. We demonstrate that the pinning of the Fermi energy to the chromium resonant level occurs for both *n*-type and *p*-type $Pb_{1-x}Sn_xTe$ in the whole composition range. This enables us to determine the valence band and conduction band offsets at the PbTe/SnTe/PbSe heterointerfaces, which is important for designing high–performance 2D transistors. Furthermore, the magnetic measurements reveal the presence of Cr ions in three charge states, $Cr^{3+}$, $Cr^{2+}$, and $Cr^{1+}$. The last one corresponds to the Cr dopants incorporated at the interstitial, and not the substitutional, sites. The measured concentrations of the interstitial and substitutional Cr are comparable.

**Keywords:** Electrical conductivity, Magnetism, Density functional calculations, IV-VI semiconductors, band offsets.

# I. INTRODUCTION

Current interest in PbTe, SnTe, and their substitutional $Pb_{1-x}Sn_xTe$ alloys is driven by their captivating physical properties, such as large relativistic effects including strong spin-orbit coupling which are reflected in the presence of the topological crystalline insulator phase, ferroelectricity, and extreme dielectric constants. On the other hand, their small band gaps (and effective dielectric screening) promoted applications in thermoelectricity and infrared optoelectronics [1–8]. Successful research and/or applications require controlled *n*- and *p*-type doping in a wide range of free carrier concentrations. In particular, very low *n* and *p* concentrations of about $10^{16}$ - $10^{17}$ cm$^3$, with negligible contribution of bulk conductivity, are necessary for investigations of topological properties. On the other hand, optimal thermoelectric properties are observed in the opposite limit, for heavy bulk *n* and *p* doping of the order $10^{19}$ - $10^{20}$ cm$^3$.

Thermoelectric properties of IV-VI compounds were enhanced thanks to the discovery that a nearly twofold increase in the thermoelectric figure of merit in PbTe can be achieved by doping with thallium [9]. The mechanism behind this effect is associated with the fact that Tl in PbTe is a resonant acceptor, whose electron level is degenerate with the valence band states. Consequently, Tl induces a pocket in the density of states (DOS) close to the valence band top, which in turn can pin the Fermi energy $E_F$ when the Tl concentration is sufficiently high. Subsequent search oriented on finding other resonant dopants and optimizing this effect included Tl [10], In [11], and transition metal (TM) ions, such as Sc [12], Fe [13], and Ni [14].

The most investigated was Cr in PbTe and $Pb_{1-x}Sn_xTe$ [13,15–28,28]. Unlike Tl, Cr in PbTe is a resonant donor, which electron level is degenerate with the conduction band states. Peculiarities of transport and magnetic properties of resonant donors in PbTe:Cr, PbSe:Cr, and HgSe:Fe semiconductors [29,30], were first compared in Ref. [16]. Information on the resonant impurities in semiconductors was summarized in two books, [2] and [31], and theoretical description of the resonant states was presented in Refs. [10,32].

Previous studies of $Pb_{1-x}Sn_xTe$:Cr demonstrated that with increasing content of Sn in the alloy the energy of the Cr level relative to the conduction band minimum (CBM) decreases. These works were limited to low concentrations of Sn in $Pb_{1-x}Sn_xTe$ (x<0.2). In this regime Cr remains a resonant donor. However, the observed trend suggests that Cr may change its character from a resonant donor to a resonant acceptor in SnTe or Sn-rich $Pb_{1-x}Sn_xTe$. This suggestion is verified in the present study.

From the experimental point of view, properties of Cr in $Pb_{1-x}Sn_xTe$ were investigated solely by electron transport measurements. The resonant character of the Cr donor was concluded based on the observed Fermi level pinning. This effect implies a mixed valence character of Cr ions in $Pb_{1-x}Sn_xTe$, i.e., a coexistence of two charge states, $Cr^{2+}$ and $Cr^{3+}$. To confirm their presence by direct observation, magnetic measurements are performed using both electron paramagnetic resonance (EPR) and magnetization measurements with an approach allowing not only to identify the actual charge states of Cr, but also to assess the corresponding concentrations [33]. We find that, in addition to the two expected $Cr^{2+}$ and $Cr^{3+}$ charge states, also $Cr^{1+}$ is present in all samples. Our density functional calculations provide a consistent interpretation of the experimental data. In particular, they show that the $Cr^{1+}$ charge state in $Pb_{1-x}Sn_xTe$ is assumed by the interstitial, $Cr_I$, and not by the substitutional $Cr_{cation}$, dopant. Their presence impacts the concentration of free carriers because the actual concentrations of substitutional and interstitial Cr ions are close. The interstitial incorporation of dopants in IV-VI compounds was not considered thus far, but it may apply to other dopants as well. Finally, the determined energies of the Cr levels in PbTe, PbSe, and SnTe allows us to estimate band offsets at interfaces between each two hosts. To this end we use the empirical approach of Langer and Heinrich [34,35], based on the observation that the energy level of a TM dopant can serve as a reference energy. Indeed, by aligning the energies of the dopant in two hosts one obtains a reasonable estimate of band discontinuities at the corresponding heterointerface. Because the offsets govern transport and optical properties of PbTe/SnTe heterostructures, they were determined in a number of experimental works [36–40]. The data exhibit a relatively large spread exceeding the host band gaps. Our results do not support the large values obtained in Ref. [40].

The paper is organized as follows. Section II presents technological details of crystal growth and morphological characterization of our samples. We also describe experimental techniques used in transport and magnetic measurements, and provide details of the first principles calculations. Since Cr doping of $Pb_{1-x}Sn_xTe$ is relatively complex given its two incorporation sites and three charge states, for the sake of clarity we provide in Sec. III a brief summary of the main theoretical results. Experimental results are presented in Section IV. They comprise transport measurements of Cr-doped PbSe and $Pb_{1-x}Sn_xTe$ alloy in the whole composition range and, in particular the temperature dependencies of carrier densities. The results are fitted using a model based on the charge neutrality equation. Next, the EPR and magnetization studies are presented. They complement the transport data providing concentrations of the three charge states of Cr. The results allow establishing energies of Cr in PbTe, SnTe and PbSe, which in turn enables us to estimate the offsets at the PbTe/SnTe, SnTe/PbSe, and PbSe/PbTe heterointerfaces. Detailed theoretical results are presented in Section V, which includes the electronic structure of both substitutional and interstitial Cr ions and their allowed charge states in PbTe and SnTe, together with the corresponding formation energies. We also consider the impact of cation vacancies on the Cr charge states. Section VI concludes the paper.

## II. SAMPLES AND METHODS
### A. Growth and morphology

Polycrystalline ingots of $Pb_{1-x-y}Sn_xCr_yTe$ are grown by the Bridgman method, covering a wide range of Sn and Cr compositions: $0 \leq x \leq 1$ and $0.005 \leq y \leq 0.04$, nominally. To ensure the conditions appropriate for the Fermi level pinning, $Pb_{1-x-y}Sn_xCr_yTe$ samples with $x$ ranging from 0.056 to 0.8 are doped with Cr in the range of $0.005 \leq y \leq 0.02$. For samples with $x \geq 0.8$, chromium is incorporated in the range $0.02 \leq y \leq 0.04$. We obtain also $Pb_{1-y}Cr_ySe$ with composition $y = 0.0035$ by the same method, thus extending our studies to selenides.

X-ray diffraction (XRD) as well as the Energy Dispersive Spectroscopy (EDS) studies of the chemical composition demonstrate Sn distribution along the ingots with segregation coefficient $k \sim 0.8$, according to Pfann's formula [41]. Moreover, the lattice parameter depends linearly on tin composition fulfilling the Vegard's law. The presence of various secondary Cr-Te phases is often observed. In order to estimate properly the amount of chromium, the Secondary Ion Mass Spectrometry (SIMS) studies as well as the Energy Dispersive X-ray Fluorescence (EDXF) analysis are performed. SIMS studies aim also at obtaining information on the uniformity of the distribution of particular components: Pb, Te, Sn and Cr. All but last elements are built-in uniformly in the crystal lattice in all investigated samples with the already mentioned increasing tin content toward the last-to freeze points. The distribution of Cr is also almost uniform for crystals with $y = 0.005$. In contrast, for Pb$_{1-x-y}$Sn$_x$Cr$_y$Te crystals with $y = 0.01$, it slightly deviates from uniformity owing to the large number of Cr-Te precipitates. Just as in the case of tin, differences in Cr content between the beginning and the end sections of the crystal ingots are observed, pointing to the segregation coefficient $k$ below 1. For further details on XRD, SEM/EDX and SIMS analyses see Sec. S.I. (A-C) in Supplemental Material [42].

### B. Experimental

After the growth, samples are checked by electron paramagnetic resonance (EPR) to identify the Cr spin and charge states in the obtained crystals [43]. These studies are performed using an X-band BRUKER EMX plus spectrometer operating at 9.5 GHz.

Magnetization is measured using Quantum Design's Superconducting Quantum Interference Device (SQUID) Magnetic Property Measurement System - MPMS-XL magnetometer with magnetic field up to 70 kOe and the Physical Property Measurement System (PPMS) with magnetic field up to 90 kOe. SQUID magnetometry is performed mainly to confirm the presence of Cr spin and charge states revealed by EPR studies and to detect Cr$^{2+}$ species, and their identification, impossible to be observed via EPR studies. It also reveals the presence of the ferromagnetic inclusions indicated in the SEM/EDS analysis and reported earlier for these compounds [24,33].

For magnetotransport studies, parallelepiped samples of dimensions about 8 x 1.5 x 0.7 mm$^3$ are cut with a 30 μm fine-wire saw and the indium electrical contacts are attached. To determine the electrical properties of the investigated samples the temperature dependencies of the resistivity tensor are measured between liquid helium and room temperatures and in the magnetic field $H$ up to 145 kOe. The measurements are performed in the low frequency AC mode using a setup consisting of four lock-in amplifiers ZI-MFLI by Zurich Instruments. The sample is powered by an internal oscillator supplied by one of the amplifiers. Simultaneously, the current and two pairs of voltages corresponding to longitudinal and transverse resistivity components are recorded.

### C. Theory

The properties of compounds (band structures, density of states, band wavefunctions) and the influence of defects are studied by *ab initio* calculations. We use the OpenMx package [44] within the local density approximation with the exchange-correlation functional proposed by Ceperly and Alder [45]. Since the involved atoms are heavy we use fully relativistic pseudopotentials provided by OpenMx. To get the proper value of the energy gap for PbTe and the proper symmetries of the valence and conduction bands we adjust spin-orbit strength for 6p(Pb) orbitals [44].

All our calculations are performed for cubic supercells containing 216 atoms. For such a supercell introduction of a single defect (chromium atom, cation vacancy) corresponds to less than 1 at.% in the bulk (i.e., to the volume concentration of about $10^{20}$ cm$^{-3}$). The distances between defects are about 19 Å, thus one may assume that the dilute limit is realized.

The convergence criterion of self-consistent calculations is $10^{-6}$ Ha and for samples the geometry optimization is performed with the force criterion $5\times10^{-4}$ Ha/Bohr. We neglect small changes of the lattice parameters caused by the defects and the experimental values are used: 6.46 Å for PbTe based systems and 6.30 Å for SnTe based systems [1] [2].

Calculations for the supercell result in folding of the Brillouin zone (BZ) which becomes much smaller than the usual BZ for fcc lattice. Four nonequivalent *L* points for 2-atom cell become a single point in the 216-atom supercell. In the following we call this point the „*L*" point. The degeneracy at the „*L*" point, including spin is 8, instead of 2 for the *L* point. Moreover, the band structures below along "*L*"-*Γ* direction contain bands along different crystallographic directions.

Finally let us notice that OpenMx provides the set of tight binding parameters which may be used to calculate density of states (DOS), partial density of states (PDOS) and the wave function structures for different systems.

### III. CHROMIUM RESONANT LEVEL IN Pb$_{1-x}$Sn$_x$Te – STATE OF THE ART

According to our study, properties of Cr in Pb$_{1-x}$Sn$_x$Te are relatively complex, and a coherent picture is obtained when the theoretical results are matched with the experimental data. Therefore, for the sake of clarity, we begin by a summary of results and conclusions provided by theoretical calculations, which constitute the conceptual framework of our interpretation of the experiment.

The calculated formation energy of the interstitial $Cr_I$ is close to that of the substitutional $Cr_{cation}$, and thus the actual concentrations of both impurities can be close as well. In ideal (i.e., defect-free) PbTe and SnTe hosts, $Cr_I$ occupying an interstitial site is a single donor and it assumes only one charge state, $Cr^{1+}$, with the $d^5$ electron configuration and spin 5/2. Its presence in our samples is confirmed in both EPR and magnetization measurements.

The properties of the substitutional $Cr_{cation}$ depend on the host. In a perfect PbTe lattice, $Cr_{Pb}$ induces a level resonant with the conduction band states. This level is an $e_g$ doublet originating from the $d$(Cr) atomic orbitals. Because of its resonant character, $Cr_{Pb}$ autoionizes and an electron occupying the $e_g$ level is transferred to the bottom of the conduction band. Hence, Cr assumes the $Cr^{3+}$ state with the $d^3$ electron configuration and spin 3/2. With increasing Cr concentration, the conduction electron concentration increases as well until the Fermi energy $E_F$ reaches the donor level of Cr. For higher Cr concentrations, $Cr^{3+}$ and $Cr^{2+}$ ions coexist in the sample, and the Fermi level pinning takes place. In the $Cr^{2+}$ charge state, the $e_g$ doublet is occupied by one electron, and the $Cr_{cation}$ spin is 2. Consequently, we consider the presence of the $Cr^{2+}$ ions as an indication of the Fermi level pinning.

Our calculations show that $Cr_{Sn}$ in SnTe is *formally* a resonant acceptor, since the $e_g$ level is degenerate with the valence band states. In the case of ideal SnTe, the Cr ion assumes the 2+ charge state, with the corresponding configuration $d^4$, spin 2, and one electron on the $e_g$ level. In principle, $Cr_{Sn}$ can attain the $Cr^{1+}$ state with a second electron on $e_g$, i.e., it can act as an acceptor. However, in this case, because of the large electron-electron repulsion between $d$(Cr) electrons, the +2/+1 transition level of Cr is situated higher in energy, in the band gap of SnTe. In consequence, we do not expect the Cr dopant to act as an acceptor increasing the hole concentration.

In the interpretation of experiment one must take into account the actual presence of cation vacancies $V_{cation}$ and the resulting high concentrations of holes. They are compensated in part by $Cr_I$ donors. The substitutional $Cr_{cation}$ donors in Pb-rich $Pb_{1-x}Sn_xTe$ also contribute to compensation. In principle, the situation in principle is less clear in Sn-rich alloys and in SnTe, because both $Cr_{cation}$ and $V_{cation}$ induce levels degenerate with the valence band. In this case, the calculated energy of $e_g(Cr_{Sn})$ is above the energy of $V_{Sn}$, one electron is transferred from $Cr_{Sn}$ to $V_{Sn}$, and compensation of vacancy-induced holes takes place. In consequence, the Cr ions are in the $Cr^{3+}$ state until the Cr concentration exceeds that of $V_{Sn}$, and pinning of the Fermi level occurs, and $Cr^{3+}$ and $Cr^{2+}$ ions coexist.

The findings above allow us to clearly discriminate between the impacts of the substitutional and interstitial Cr ions. Indeed, because the spin states of $Cr_I$ and $Cr_{cation}$ are different, magnetic measurements can distinguish the two incorporation sites of chromium as well as provide the corresponding concentrations. Transport measurements provide, in turn, the concentration of free carriers, the Fermi energy, and energies of Cr states.

Stabilization of the position of the Fermi level in PbTe can be explained in terms of the energy diagrams presented in Fig. 1. $Cr^{2+}$ ions are expected to substitute $Pb^{2+}$ ions in PbTe crystal lattice. Electrons originating from the self-ionization process of the $Cr^{2+}$ ions ($Cr^{2+} \rightarrow Cr^{3+} + e^-$) primarily fill the empty electron states in the valence band originated mainly from the metal vacancies and, subsequently, also the electron states in the conduction band (Fig. 1(a)). More specific discussion related to charge states of chromium is presented in section V, Theory. The change of the type of conductivity with increasing Cr content is an interplay between chromium donors and the density of metal vacancies. As a result, samples with small Cr content can still be p-type. Electron concentration saturates when the Fermi level reaches the $Cr^{2+/3+}$ energy level. Consequently, self-ionization process stops as otherwise it would result in an energy increase of the electronic system (Fig. 1(b)). Further increase of Cr content does not lead to further rise of the electron concentration. However, by co-doping PbTe crystals with other donor impurities, for instance iodine, we may increase the Fermi energy, which results in obtaining chromium in the $Cr^{2+}$ state solely (Fig. 1(c)).

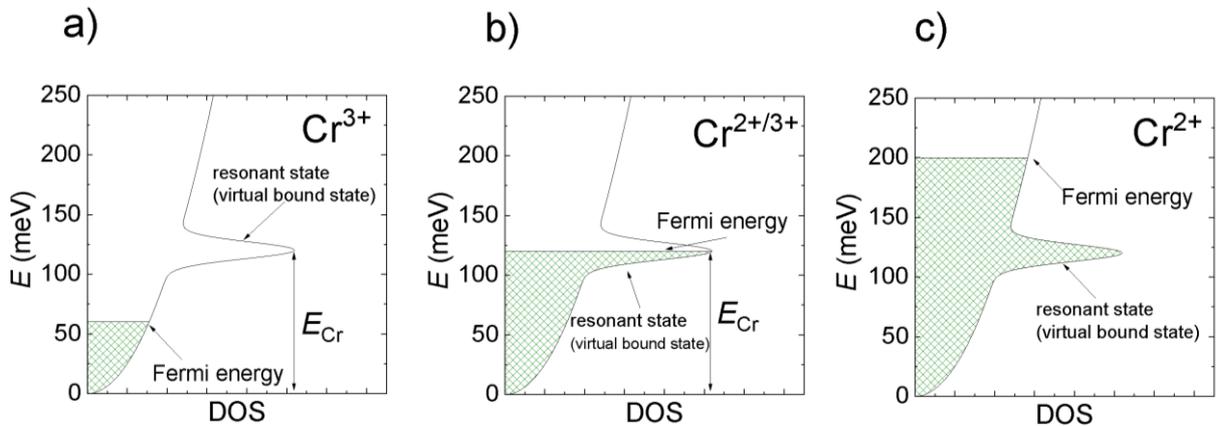

FIG. 1. Schematic model of the energy, $E$, dependence on the density of states, DOS, in the conduction band of $Pb_{1-y}Cr_yTe$. Lorentzian shaped DOS contribution superimposed on the conduction band represents resonant $Cr^{2+/3+}$ level, located at energy $E_{Cr}$ above the conduction band minimum. (a) The Fermi energy ($E_F$) lies below the $Cr^{2+/3+}$ level. There is no $E_F$ pinning and all Cr ions are in the $Cr^{3+}$ state. (b) Cr content exceeds $y = 0.001$ and $E_F$ approaches the $Cr^{2+/3+}$ resonant level: both - $Cr^{3+}$ and $Cr^{2+}$ charge states coexist. (c) Conduction band is filled above $Cr^{2+/3+}$ level by co-doping with iodine. $Cr^{2+}$ ions do not undergo self-ionization.

Here we present the $E_F$ behavior in the presence of the resonant Cr level for the $Pb_{1-y}Cr_yTe$ system. However, we emphasize that this model is as well valid for Cr doped SnTe and PbSe, to be discussed in next sections.

## IV. RESULTS

### A. Electron transport measurements

Electron concentration $n$ as a function of Cr content $y$ is shown for both $Pb_{1-y}Cr_yTe$ and $Pb_{1-y}Cr_ySe$ in Fig. 2. The initial increase of $n$ with increasing Cr content saturates at a certain value of $y$, i.e., a further increase of $y$ does not affect the electron concentration, and the Fermi energy $E_F$ is constant. This is the experimental confirmation of the concept shown schematically in Fig.1. In the case of $Pb_{1-y}Cr_yTe$, the carrier concentration in the saturation region is $n_{sat}(PbTe)=1.2\times10^{19}$ cm$^{-3}$, which corresponds to the Fermi energy $E_F(PbTe)=0.10$ eV, whereas for $Pb_{1-y}Cr_ySe$ we find $n_{sat}(PbSe) = 1.8\times10^{19}$ cm$^{-3}$ and $E_F(PbSe) = 0.12$ eV. Details of the band structure calculations are given in Sec. S.II. of the Supplemental Material.

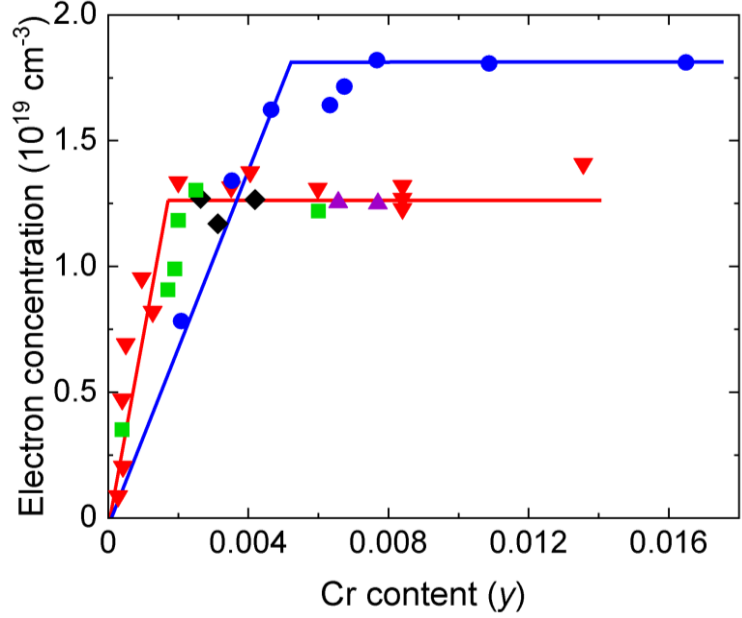

FIG. 2. Electron concentration versus chromium content, $y$, in $Pb_{1-y}Cr_yTe$ and $Pb_{1-y}Cr_ySe$ measured at 4.2 K. Red inverted triangles and blue circles are the results for $Pb_{1-y}Cr_yTe$ and $Pb_{1-y}Cr_ySe$, respectively, from Ref. [26], black diamonds are results for $Pb_{1-y}Cr_yTe$ from Ref. [27], violet triangles – data for $Pb_{1-y}Cr_yTe$ from Ref. [28], green squares – presently reported data for $Pb_{1-y}Cr_yTe$. Solid lines are to guide the eyes.

The hole concentration versus Cr content $y$ in SnTe samples is presented in Fig. 3. Starting from concentration $p = 2.2 \times 10^{21}$ cm$^{-3}$ ($y = 0$) the density of holes systematically decreases with increasing Cr content. This indicates that Cr in SnTe acts as a donor, like in PbTe. Assuming that Cr is a single donor, we calculate the reduction of $p$ with the composition $y$, and plot it in Fig. 3 as a red line. Despite the high spread of hole concentrations we find these results reasonable due to high variety of $p$-type background concentration in pure SnTe as a consequence of different crystal growth conditions.

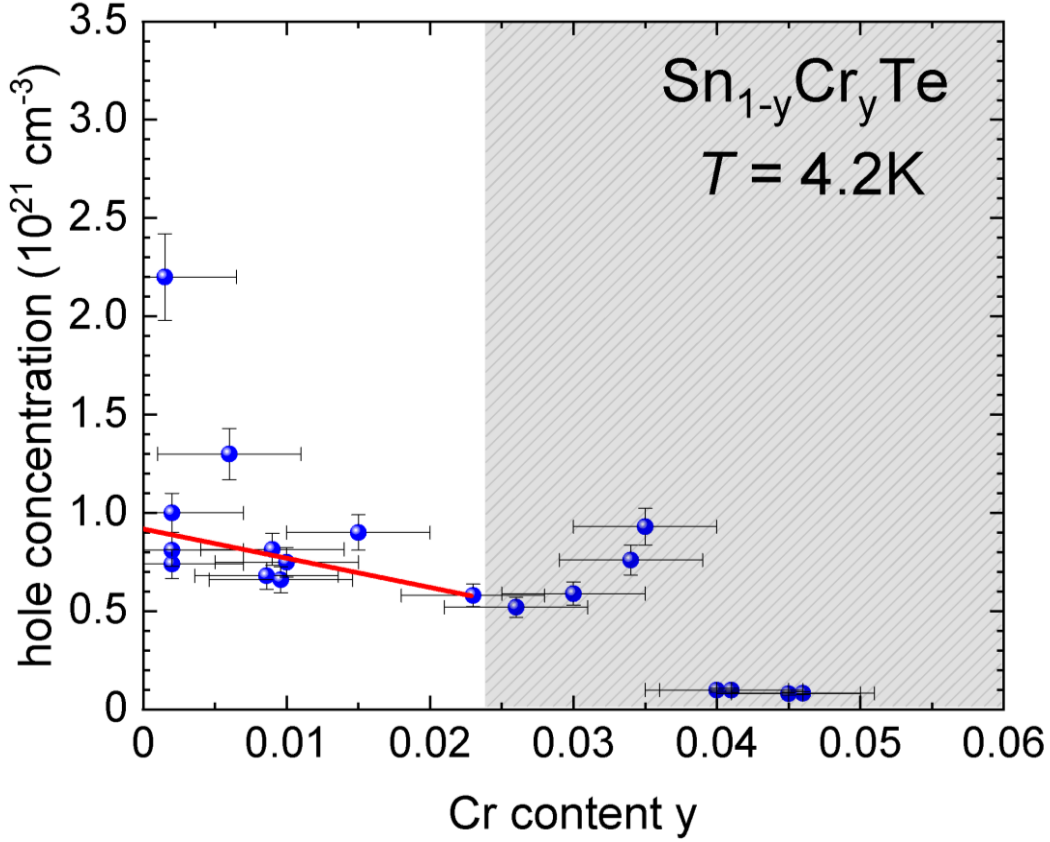

FIG. 3 Hole concentration versus chromium content, $y$, in $Sn_{1-y}Cr_yTe$ obtained at 4.2 K. Blue spheres are the experimental results, and the red solid line describes the impact of Cr donors on the hole concentration, see text. Samples with $y > 0.02$ contain Cr-Te inclusions, therefore they are marked with grey dashed area.

Important information regarding the energy of the Cr ions in the whole composition range of $Pb_{1-x-y}Sn_xCr_yTe$ is provided by the measurements of the temperature dependencies of the Hall coefficient. The measured electron concentrations as a function of temperature for selected PbTe samples with varying chromium concentration are shown in Fig. 4. Analogous dependencies are obtained for several $Pb_{1-x-y}Sn_xCr_yTe$ samples with $0 < x < 0.17$ presented in Fig. 5. As can be seen, the electron concentration is initially independent of temperature and than it drops rapidly at a certain onset temperature. In both figures, the data are described with good accuracy by our model of the $Pb_{1-x-y}Sn_xCr_yTe$ band structure, outlined briefly below.

We apply the two-band $\mathbf{k} \cdot \mathbf{p}$ model with nonparabolic energy dispersion relation [46]:

$$E = -\frac{E_{gap}}{2} + \left[\left(\frac{E_{gap}}{2}\right)^2 + \frac{E_{gap}\hbar^2 k^2}{2m^*}\right]^{1/2} \quad (1)$$

where $E_{gap}$ – energy gap, $\hbar$ – Plack constant, $k$ – wave vector, $m^*$ - the density of states effective mass. Both conduction and valence bands are nonspherical with the bands' minima located at four equivalent $L$ points of the BZ. For the sake of simplicity of carrier density calculations, we assume spherical energy bands with the above displayed nonparabolicity.

Band structure parameters such as the energy gap $E_{gap}$ and the density of states effective masses $m^*$ are taken from the experimental data published in Refs. [1,47]. The temperature dependence of the effective mass is modeled with a 3rd order polynomial and the values of the coefficients are listed in the Supplemental Material (Table S3). The details of the calculations, together with the appropriate equation for the energy gap versus temperature and composition (using Varshni formula [47]) are given in Sec SII of Supplemental Material.

We calculate the Fermi energy by iteratively solving the simplified neutrality equation, namely the electron concentration in the conduction band is equated to the number of ionized Cr ions. We neglect here the contributions from intrinsic electron and hole densities (almost two orders of magnitude lower at room temperture than the real observed carrier concentrations of the material) as well as the $p$-type background related to the native metal vacancies. The resonant density of Cr states is modeled by the Lorentzian shape similarly to Fe resonant impurity in HgSe studied in Refs. [48–50]. As a result we obtain two important $T$ and $x$ independent parameters necessary for the description of all experimental data: the $Cr^{2+/3+}$ level position in pure PbTe $E_{res}$ (0 K) = 95 meV and its width $W_{res}$ = 3 meV. We point out here, moreover, that the $T$ and $x$ dependence of the $Cr^{2+/3+}$ level is not another fitting parameter but results from the band-gap changes. Further details and all the necessary formulas are presented in the Supplemental Material (Sec. S.II.).

As shown in Figs. 4 and 5, the model quite accurately accounts for the experimental data. The observed decrease of the electron concentration results from the temperature dependence of the energy gap and the change of the $Cr^{2+/3+}$ level energy with addition of tin.

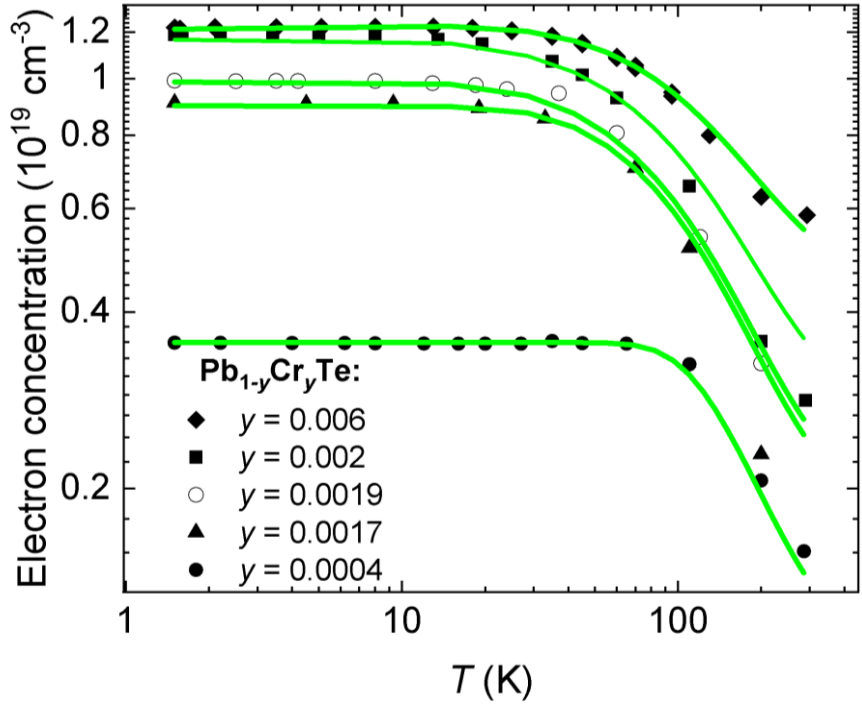

FIG. 4. Temperature, $T$, variation of the electron concentration for various Cr content, $0.0004 \leq y \leq 0.006$, in $Pb_{1-y}Cr_yTe$ samples. Symbols denote experimental results and green solid lines represent solution of the neutrality equation model.

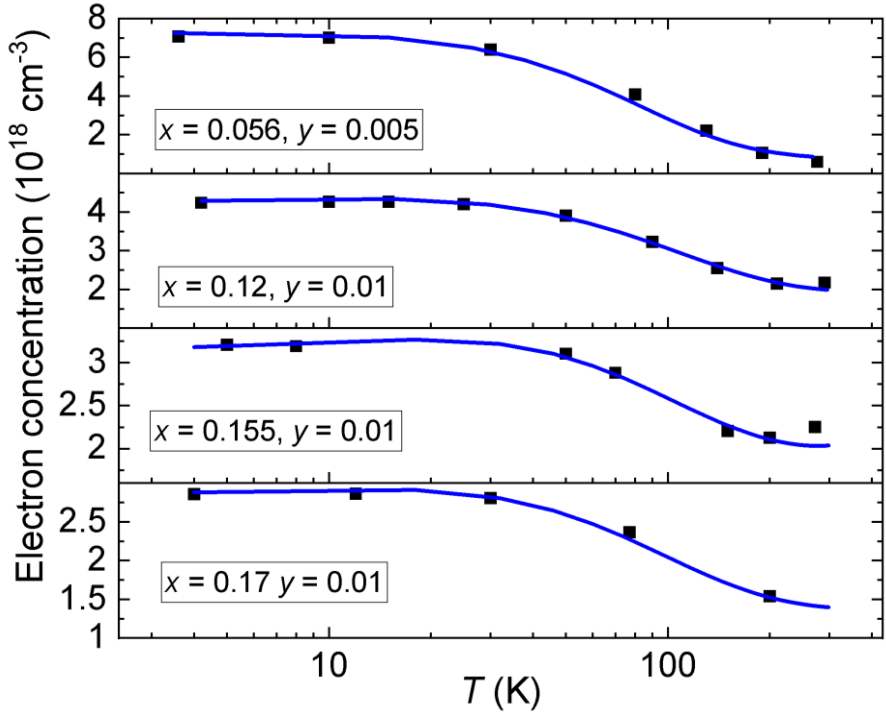

FIG. 5. Temperature, $T$, variation of the electron concentration for several $Pb_{1-x-y}Sn_xCr_yTe$ samples, $0.056 \leq x \leq 0.17$ and $y = 0.01$. Black squares denote experimental data and blue solid lines represent the solution of the neutrality equation model.

## B. Identification of Cr spin and charge states via magnetic measurements

In order to support our interpretation of the transport results, in this Section, we present and discuss in this Section the experimental data on spin and charge states of isolated paramagnetic Cr ions in $Pb_{1-x-y}Sn_xCr_yTe$ in the entire Sn composition range ($0 \leq x \leq 1$), acquired through EPR spectroscopy and magnetometry.

We begin by discussing the results obtained via EPR spectroscopy. Representative EPR spectra collected at 3 K are presented in Fig. 6 for Sn contents, x, given in the figure and a nominal Cr content of y=0.005. In the figure, only the spectra of samples which did not show ferromagnetic signatures of Cr-Te precipitates are depicted. Two asymmetric, Dysonian-shaped signals characteristic for conducting samples are observed depending on Sn content, identified as substitutional $Cr^{3+}$, with a donor-like effective g-factor of 1.94 [15], and interstitial $Cr^{1+}$ (see Secs. III and V) with an acceptor-like g-factor of 2.05. The $Cr^{3+}$ spin state is observed only at low Sn contents, and its EPR signal vanishes above $x = 0.12$. The $Cr^{2+}$ charge state cannot be detected in EPR owing to large zero-field splitting of the spin quituplet ground state, which exceeds by far the transition energies available in an X-band spectrometer. As shown in Fig. 6, when $x = 0$ (PbTe) only the EPR signal of $Cr^{3+}$ ions is detected while for $x = 1$ (SnTe) only the signal of interstitial $Cr^{1+}$ is seen. On the other hand, in a number of samples with intermediate Sn content we detect the EPR signals of both chromium in the $Cr^{1+}$ and $Cr^{3+}$ charge states. We note, that in such samples the integrated intensity of the interstitial $Cr^{1+}$ EPR signal cosiderably exceeds that of the substitutional $Cr^{3+}$.

Finally, an additional, narrow-line EPR signal with a donor-like *g*-factor of 1.997 is observed in almost all samples (see Fig. 6). The same donor signal was previously reported in polycrystalline PbTe doped with Cr [15]. Since the signal occurs both in *n*-type and *p*-type samples we attribute it, as earlier, to defect on the grain boundaries.

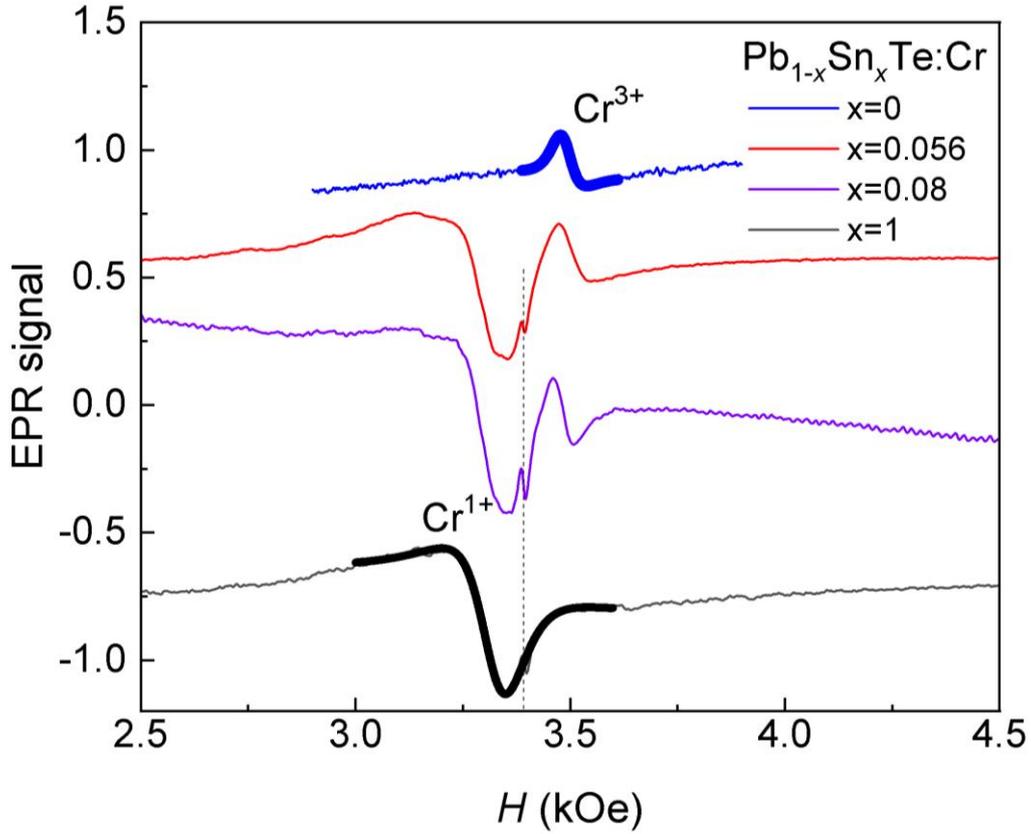

FIG. 6. X-band electron paramagnetic resonance spectra (magnetic field, $H$, derivative of the absorbed microwave power) for selected $Pb_{1-x-y}Sn_xCr_yTe$ samples, measured at 3 K. Thick solid lines represent the Dysonian line fits with corresponding $g$-factors for the two observed charge states ($Cr^{3+}$ and $Cr^{1+}$) collected in Table I. . The $g$-factors depend slightly on carrier concentration due to the Knight shift. The verticical dashed line denotes the position of the donor signal related to the a defect on the grain boundaries.

To put an absolute bar on the EPR findings, we resorted to direct low-temperature magnetometry. The central experimental difficulty is that our crystals invariably contain nanoscale ferromagnetic Cr–Te precipitates ($Cr_mTe_n$), whose ordering temperatures exceed 50 K (see Sec. S.I. (B) and (C) in the Supplemental Material). At the base-temperature window we study (2 - 8 K), their moment is therefore already saturated and essentially independent of both field and temperature, whereas the signal from isolated Cr ions remains strongly temperature-dependent. This disparity allows us to separate the two contributions quantitatively. Our procedure, previously validated for $Pb_{1-y}Cr_yTe$ [33], is a differential method. The core idea is that for many systems composed of multiple magnetic components, only the paramagnetic contribution varies appreciably with temperature in the low-temperature regime. If the field- and temperature-dependence of other terms - such as diamagnetism of the lattice, ferromagnetic precipitates, or blocked superparamagnetic entities - is weak by comparison, then the difference between two magnetization isotherms, $M(H,T)$, measured at nearby low temperatures, effectively isolates the change in the paramagnetic signal [51]. This differential approach has been successfully used to separate the PM component from blocked superparamagnetic or FM signals in a variety of systems, including diluted-magnetic-semiconductor films [52], bulk crystals [33,53], and nanowires [54,55], providing quantitative information on these components at the same time.

In our case, magnetization isotherms $M(H,T)$ are recorded at three temperatures (2 K, 5 K, and 8 K). For each sample we form two differences: $\Delta M_{2-5}(H) = M(H,2K) - M(H,5K)$ and $\Delta M_{5-8}(H) = M(H,5K) - M(H,8K)$, which cancel the temperature-invariant ferromagnetic term and leave only the change in the paramagnetic response.

The residual signal is modeled as a linear combination of the differences of Brillouin functions ($B_S$) calculated for the three charge states observed in our system - $Cr^{3+}$ ($S = 3/2$), $Cr^{2+}$ ($S = 2$), and $Cr^{1+}$ ($S = 5/2$):

$$\Delta B_S(H, \Delta T) = B_S(H, T_1) - B_S(H, T_2) \qquad (2)$$

Because $\Delta B_S$ exhibits a well-defined maximum whose position shifts systematically with the spin quantum number $S$, simultaneous fitting of the three functions $\Delta B_{3/2}$, $\Delta B_2$, and $\Delta B_{5/2}$ to $\Delta M_{2-5}(H)$ and $\Delta M_{5-8}(H)$ (with $g$-factors listed in Table I) yields the concentrations [Cr3+], [Cr2+], and [Cr1+], respectively, with good numerical stability. Introducing the third component, $Cr^{1+}$, extends the earlier two-state analysis [33] and refines the fit, as the two independent $\Delta B_S(H, \Delta T)$ datasets are constrained by a single set of concentrations.

TABLE I. Values of spin and g-factor for different chromium charge and spin states used in our analysis. The departure of g-factors measured in our EPR studies from $g = 2.00$ for $Cr^{3+}$ and $Cr^{1+}$ result from spin-orbit coupling and crystal field splitting according to Ref. [56], as well as from the carrier concentration dependent Knight shift (Ref. [57]). As $Cr^{2+}$ is not detected in EPR, we assume $S = 2$ corresponding to quenched orbital effects [58].

| Cr charge state | Spin | g-factor |
| --- | --- | --- |
| $Cr^{3+}$ | 1.5 | 1.94 |
| $Cr^{2+}$ | 2.0 | 2.00 |
| $Cr^{1+}$ | 2.5 | 2.05 |

Figures 7–9 illustrate the procedure for Cr-doped $Pb_{1-x-y}Sn_xTe$, PbSe, and PbTe:Cr,I samples. The excellent agreement between experiment and theory confirms the robustness of the method and, in particular, verifies the non-negligible population of $Cr^{2+}$ ions (Fig. 8). This approach offers a rare experimental means to distinguish and quantify multiple Cr charge states in a bulk crystal - something typically inaccessible to conventional magnetometry or EPR - and is therefore crucial for understanding and modeling the Fermi-level pinning mechanism and its role in stabilizing the band structure across a wide composition range. The presence of the donor-like $Cr^{2+}$ state across the entire Sn-composition range is a direct experimental signature of Fermi-level pinning by the resonant Cr level in $Pb_{1-x-y}Sn_xCr_yTe$.

Each panel in Fig. 7(b-i) is subject to the same conditions, schematically illustrated in Fig. 7(a). Namely, since Fermi energy is located below (in the conduction band) or above (in the valence band) the $Cr^{2+/3+}$ level, all the chromium should be in $Cr^{3+}$ charge state, as mentioned in Sec. III (Fig. 1 (a)). The resulting concentrations of chromium in particular charge states are displayed in each panel.

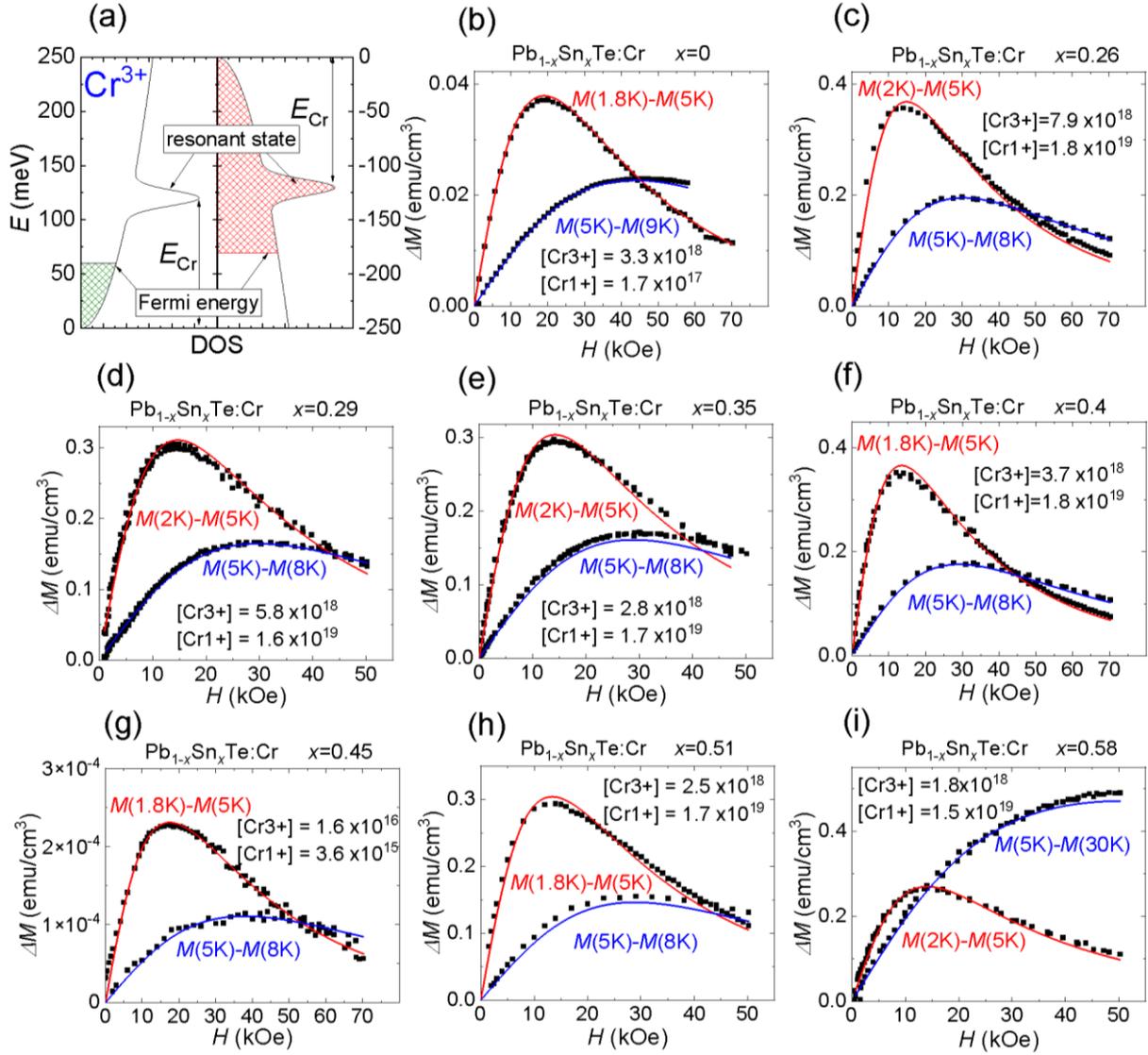

FIG. 7. Magnetic field, $H$, dependence of magnetization differences, $\Delta M$ taken between two adjacent temperatures for $Pb_{1-x-y}Sn_xCr_yTe$ samples with varying Sn compositions, (a) Energy, $E$, dependence of the density of states, DOS, for PbTe (left panel) and SnTe (right panel) with low chromium contents. The location of the Fermi level $E_F$ below the resonant $Cr^{2+/3+}$ level is indicated. (b-i) Results for $Pb_{1-x-y}Sn_xCr_yTe$ with (b) $x = 0$, (c) $x = 0.26$, (d) $x = 0.29$, (e) $x = 0.35$, (f) $x = 0.40$, (g) $x = 0.45$, (h) $x = 0.51$, and (i) $x = 0.58$. Symbols denote experimental data. Solid curves are the results of magnetization data fitting in accordance with equation 4 taken for three spin values of Cr: $S = 3/2$, $S = 2$, $S = 5/2$. The obtained concentrations of magnetic ions in particular charge states (symbols in square brackets) are given in cm$^{-3}$.

Figure 8(b - k) shows implementation of the same approach but for $Pb_{1-x-y}Sn_xCr_yTe$ samples exhibiting Fermi level pinning effect, which is schematically shown in Fig. 8(a). In each sample in this set Cr occurs in three charge states: 3+, 2+, and 1+. For better insight into the coexistence of various charge states of Cr we perform decomposition of the magnetization

data for Pb$_{1-x-y}$Sn$_x$Cr$_y$Te samples from Fig. 8 into particular paramagnetic components (Fig. S6 in sec. S.III. of Supplemental Material). As shown in Fig. 8 the concentrations of Cr ions in particular charge state - and most importantly that of Cr$^{3+}$ - differ from sample to sample, which may seem surprising. . This is related to changes in the inherent background hole concentration of Pb$_{1-x}$Sn$_x$Te depending on different growth conditions, as explained later (Fig. 10).

We also provide the same fitting analysis of fitting performed for Cr doped PbSe in Fig. 8 (l). The presence of Cr$^{2+}$ ions additionally confirms the Fermi level pinning in the studied system, moreover in the case of PbSe this occurs at higher energies as compared to PbTe, which is also reflected in Fig. 2 for several other Cr doped PbSe samples.

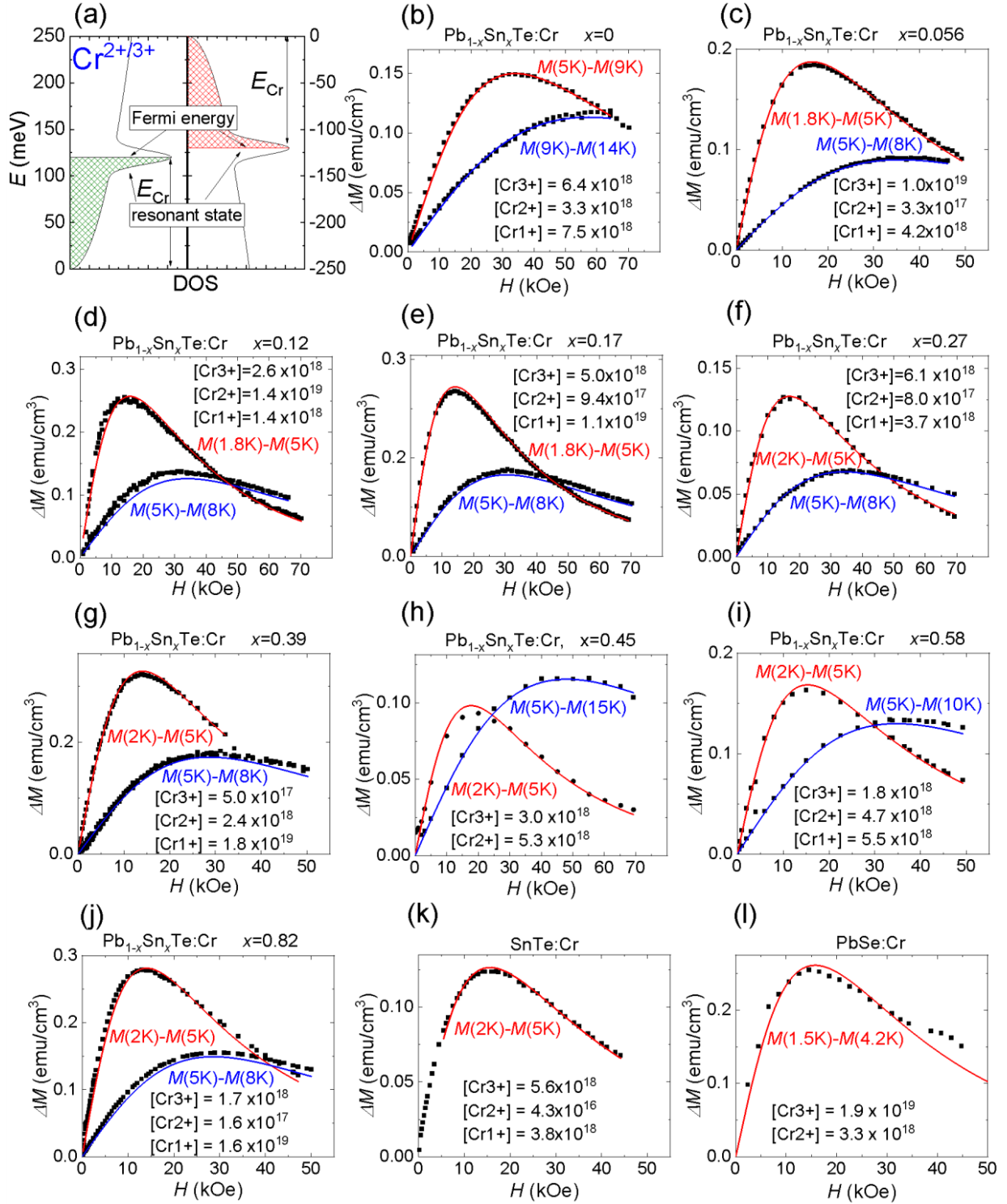

FIG. 8. Magnetic field, $H$, dependence of magnetization differences, $\Delta M$, taken between two adjacent temperatures for $Pb_{1-x-y}Sn_xCr_yTe$ samples with varying Sn compositions $0 \leq x \leq 1$ and $Pb_{1-y}Cr_ySe$. (a) Energy, $E$, dependence of the density of states, DOS, for PbTe (left panel) and SnTe (right panel). The Fermi energy $E_F$ lies within the enhanced DOS of the resonant $Cr^{2+/3+}$ states. (b) $x = 0$, (c) $x = 0.056$, (d) $x = 0.0.12$, (e) $x = 0.17$, (f) $x = 0.27$, (g) $x = 0.39$, (h) $x = 0.45$, (i) $x = 0.58$, (j) $x = 0.82$, and (k) $x = 1$. (l) $Pb_{1-y}Cr_ySe$. Symbols denote experimental data and solid curves are the results of magnetization data fitting. The obtained

concentrations of magnetic ions in particular charge states (symbols in square brackets) are given in cm$^{-3}$.

To complete the description of possible locations of the Fermi level mentioned in Sec.III (see Fig. 1 (c)), we present another system – Pb$_{1-y}$Cr$_y$Te codoped with iodine, which is a regular nonmagnetic and non-resonant donor. Doping with iodine enables us to obtain higher electron concentration, significantly exceeding the pinning limit, which consequently results in the absence of chromium in Cr$^{3+}$ charge state in this sample (see Fig. 9).

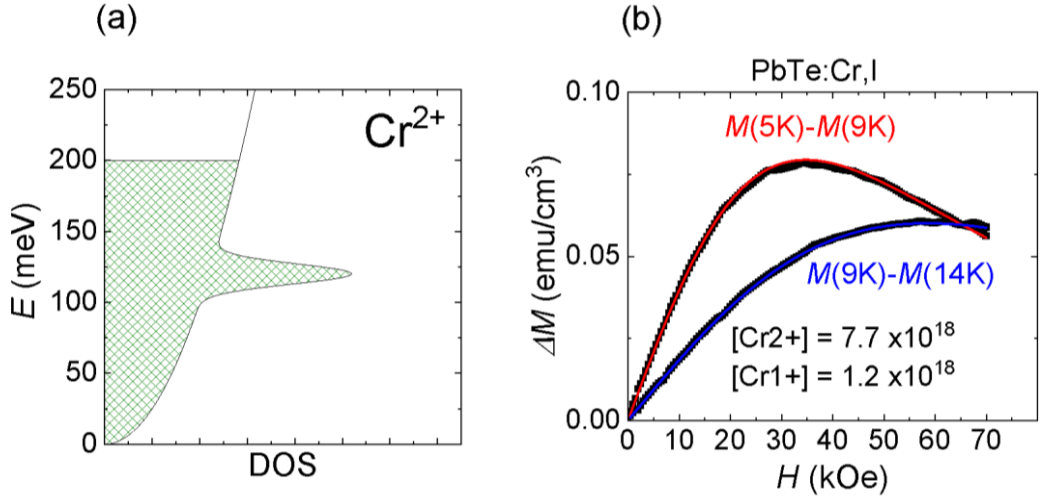

FIG. 9 (a) Energy, $E$, dependence of the density of states, DOS, in PbTe:Cr codoped with iodine. The Fermi energy is located above the enhanced DOS of the resonant Cr$^{2+/3+}$ state. (b) Magnetic field, $H$, dependence of magnetization differences, $\Delta M$, taken between two adjacent temperatures in PbTe codoped with Cr and I. The obtained concentrations of Cr in 2+ and 3+ charge states (symbols in square brackets) are given in cm$^{-3}$.

This final test provides the complete description of Cr$^{2+/3+}$ states in the whole range of tin composition in Pb$_{1-x-y}$Sn$_x$Cr$_y$Te system, as well as SnTe and PbSe doped with chromium. The coexistence of Cr$^{2+}$ and Cr$^{3+}$ charge states in both $n$ and $p$-type samples ($0 \leq x \leq 1$) indicates that knowing the particular position of the Fermi energy we determine precisely the position of the Cr$^{2+/3+}$ level in the IV - VI ternary compounds.

The concentrations of Cr in Pb$_{1-x-y}$Sn$_x$Cr$_y$Te in each charge state are shown in Fig. 10 together with the Hall concentrations of free carriers. Figs 7 and 8 show that Cr ions in the Cr$^{1+}$ charge state are present in all samples. In addition, our samples can be split into two sets: those which do not contain Cr$^{2+}$ ions are displayed in Fig. 7, and those containing both Cr$^{3+}$ and Cr$^{2+}$ ions are displayed in Fig.8.

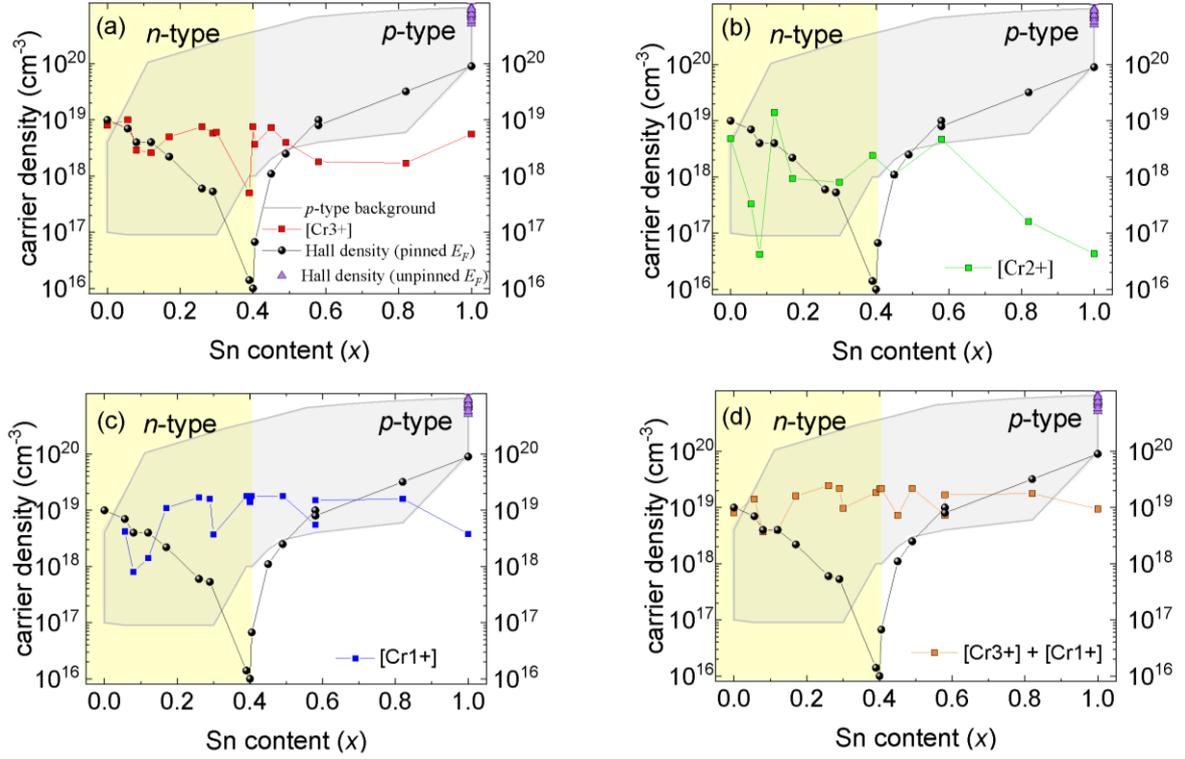

FIG. 10. Concentrations of Cr ions (squares) and Hall concentrations of free carriers (black dots) as a function of the alloy composition. The Cr charge state is specified in square brackets. (a) [Cr3+], (b) [Cr2+], (c) [Cr1+], and (d) the sum of [Cr1+] and [Cr3+]. The left sides of each panel marked in yellow indicate the composition window $x < 0.4$ in which $Pb_{1-x}Sn_xTe$ is of $n$-type, while for $x > 0.4$ on the right sides of the panel $Pb_{1-x}Sn_xTe$ is $p$-type. The theoretical concentration range of cation vacancies is contained within the area limited by gray solid lines. Red triangles indicate the highest hole concentration in $Sn_{1-y}Cr_yTe$ samples, confirming the thermodynamic boundaries for concentrations [2] indicated by the gray lines.

Anticipating the results of our calculations (see Sec. V Theory) we point out that there are two incorporation channels of single Cr ions, the interstitial and the substitutional one. In the former case, the interstitial $Cr_I$ does not provide electrons for bonding and assumes only the $Cr^{1+}$ charge state (corresponding to the $d^5$ configuration) in both PbTe and SnTe, and it is a single donor. Typically, its concentration assumes the level of about $10^{19}$ cm$^{-3}$ independent of the alloy composition, Fig. 10 (c), which may be considered as a solubility limit of this defect. On the other hand, the character of the substitutional Cr ion, $Cr_{sub}$, depends on the matrix: it is a resonant donor in PbTe, and it passivates holes in SnTe. $Cr_{sub}$ can assume two charge states, and the proportions of the concentrations of $Cr^{3+}$, [$Cr^{3+}$], to $Cr^{2+}$, [$Cr^{2+}$], vary from sample to sample. In the samples shown in Fig. 7, $Cr^{3+}$ is obviously the dominant charge state because $Cr^{2+}$ is absent. However, even in the samples from Fig. 8, the concentration of the ionized $Cr^{3+}$ typically exceeds that of the neutral $Cr^{2+}$.

Figure 10 also includes the measured Hall concentrations of free carriers in the analized samples. The relation between Cr dopants and free carriers is not direct because of the presence of cation vacancies, the dominant native defects in $Pb_{1-x}Sn_xTe$. They are resonant double-acceptors, with an acceptor level well below VBM, and typically occurr at concentrations of about $1 \times 10^{17}$ - $4 \times 10^{18}$ cm$^{-3}$ in PbTe, and $9 \times 10^{19}$ – $1 \times 10^{21}$ cm$^{-3}$ in SnTe. The theoretical window of vacancy concentrations calculated in Ref. [2] is shown by gray lines in Fig.10 within the whole composition range of $Pb_{1-x}Sn_xTe$. In particular, theoretical results agree well with our experimental data since in SnTe the observed hole concentrations reach $10^{21}$ cm$^{-3}$. Unfortunately, we have no tools to determine their actual concentrations. Finally, cation vacancies are the only native acceptors in $Pb_{1-x}Sn_xTe$ [1,2].

Summarizing, let us conclude the section with the following remarks:

**PbTe**. Several samples with increasing Cr content are presented in Sec. III, where the saturation of free electrons is analized in detail. In the sample with $x = 0$, reported in Fig. 10, the electron concentration is $10^{19}$ cm$^{-3}$, which agrees well with the concentrations of the two donors, [Cr1+] and [Cr3+], while that of the neutral Cr [Cr2+] = $5 \times 10^{18}$ cm$^{-3}$. These results indicate that the cation vacancies do not play an important role, which is plausible because their typical concentrations are of the order of $10^{18}$ cm$^{-3}$.

**SnTe**. The total concentration of donors, [Cr1+] and [Cr3+], amounts to $10^{19}$ cm$^{-3}$, which is an order of magnitude lower than the hole concentration $p = 10^{20}$ cm$^{-3}$, and the dominant acceptors are $V_{cat}$. The neutral $Cr^{2+}$ ions are detected at a very low level in Sn-rich samples with $x = 1.0$ and $x = 0.8$.

**$Pb_{1-x}Sn_xTe$**. With increasing Sn content $x$, the electron concentration decreases, vanishing at about $x = 0.4$. For higher $x$, the conductivity changes from $n$- to $p$-type, which agrees with the increasing concentration of vacancies. The three samples with $x \approx 0.4$ reported in Fig. 10 represent an important case, since the concentrations of both conduction electrons and holes almost vanish (i.e., they are close to $2 \times 10^{16}$ cm$^{-3}$). This implies an almost complete compensation of vacancies by Cr donors. We also note that at this composition another effect, namely the band inversion, takes place. Consequently, the Fermi level $E_F$ is in the middle of the conduction and valence band inversion region. In these samples all three charge states of Cr coexist. The concentrations of ionized donors, [Cr1+] = $2\times10^{19}$ cm$^{-3}$ and [Cr3+] = $8\times10^{18}$ cm$^{-3}$ imply that [$V_{cat}$] ≈ $2.5\times10^{19}$ cm$^{-3}$, which is reasonable. More importantly, a relatively high [Cr2+] = $2\times10^{18}$ cm$^{-3}$ provides a hint about the coincidence of both Cr level and $E_F$ levels.

### C. Determination of the PbTe/SnTe/PbSe band offsets using Cr$^{2+/3+}$ level

In this Section we evaluate the band discontinuities at the PbTe/SnTe/PbSe interfaces based on our experimental data. To this end, we follow the approach proposed in Refs. [34,35], and employ the Cr level as the reference energy serving to align the CBM and VBM at the considered interfaces. The results obtained for III-V and II-VI semiconductor heterostructures using various TM dopants [34,35] confirm the efficiency of this approach. We also note that its accuracy is limited by the fact that several factors that determine the actual offsets are neglected, e.g., formation of interface dipoles, the interface orientation and the presence of strain. In particular, in the rock salt structure the most frequently used (111)-oriented interfaces are polar. This orientation implies the presence of macroscopic electric fields in heterostructures which affect the spacial distribution of carriers. Next, energy bands are affected by misfit strains induced by the difference in lattice constants of e.g. PbTe and SnTe, and also by lattice mismatched substrates. For these reasons, a direct comparison of our results with experiment should be taken with care.

When considering band discontinuities at the PbTe/SnTe or the PbSe/SnTe interface one must take into account the fact that the band gap of SnTe is "inverted" relative to that of PbTe and PbSe. Accordingly, we define band offsets for the $L_6^+$ and $L_6^-$ bands as

$$\Delta E(+) = E_{SnTe}(L_6^+) - E_{PbTe}(L_6^+) \quad \text{and} \quad \Delta E(-) = E_{SnTe}(L_6^-) - E_{PbTe}(L_6^-). \quad (3)$$

This definition is non-ambiguous in the considered case. However, to follow the common convention we also use the definition of the valence band VBO and conduction band CBO offset in a A/B heterostructure as:

$$VBO = E_{VBM}(A) - E_{VBM}(B) \qquad (4)$$

and

$$CBO = E_{CBM}(A) - E_{CBM}(B) \qquad (5)$$

where $E_{VBM}$ and $E_{CBM}$ are the energies of the VBM and CBM, respectively.

Figure 11 presents the results of our Hall measurements together with the literature data from Ref. [25]. High Cr amount in the crystals with $x < 0.4$ ($0.01 \leq y \leq 0.02$) ensures that all samples before the band-crossing point are of *n*-type. For higher alloy compositions the conductivity changes to *p* type. We stress that none of our electric measurements results displays magnetic hysteresis loops (visible in some part of our magnetization plots), which indicates that ferromagnetic inclusions do not contribute to electron transport.

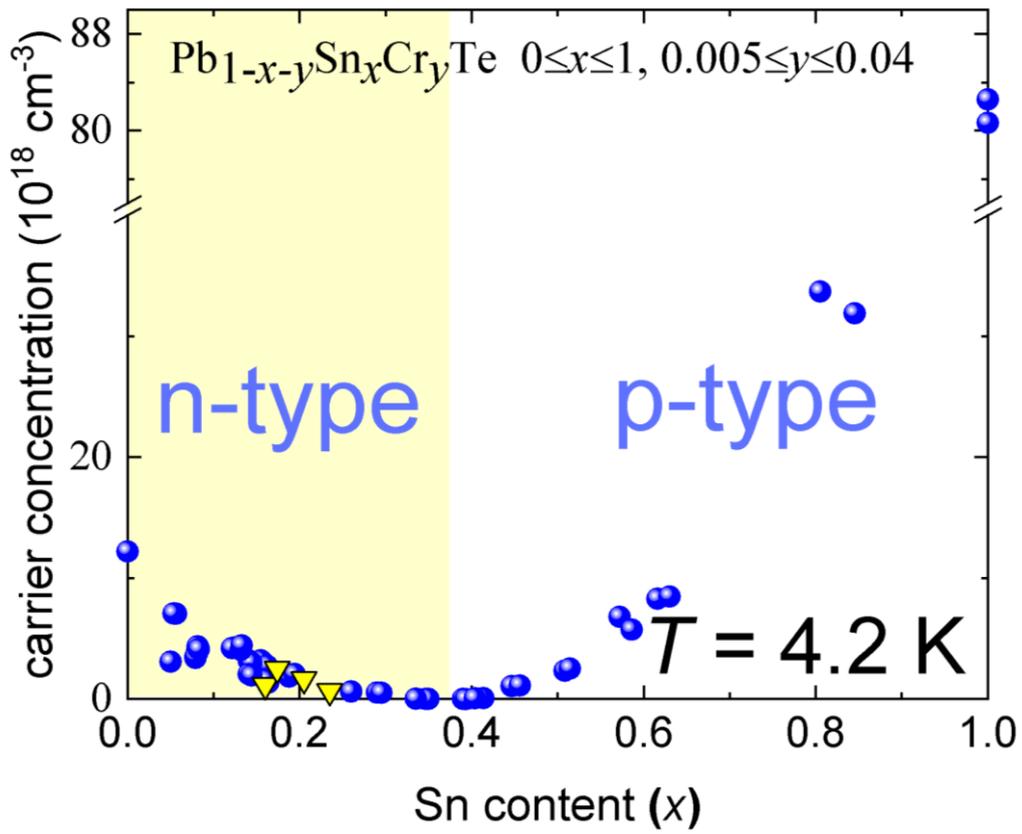

FIG. 11. Carrier concentration versus Sn content $x$ in $Pb_{1-x-y}Sn_xCr_yTe$ measured at $T = 4.2$ K. The plot includes only the samples with carrier densities high enough to observe the Fermi level pinning. Blue spheres represent our experimental data, yellow triangles - experimental results taken from Ref. [25]. The region shaded in yellow denotes the samples with $n$-type conductivity, while the white region corresponds to the $p$-type samples.

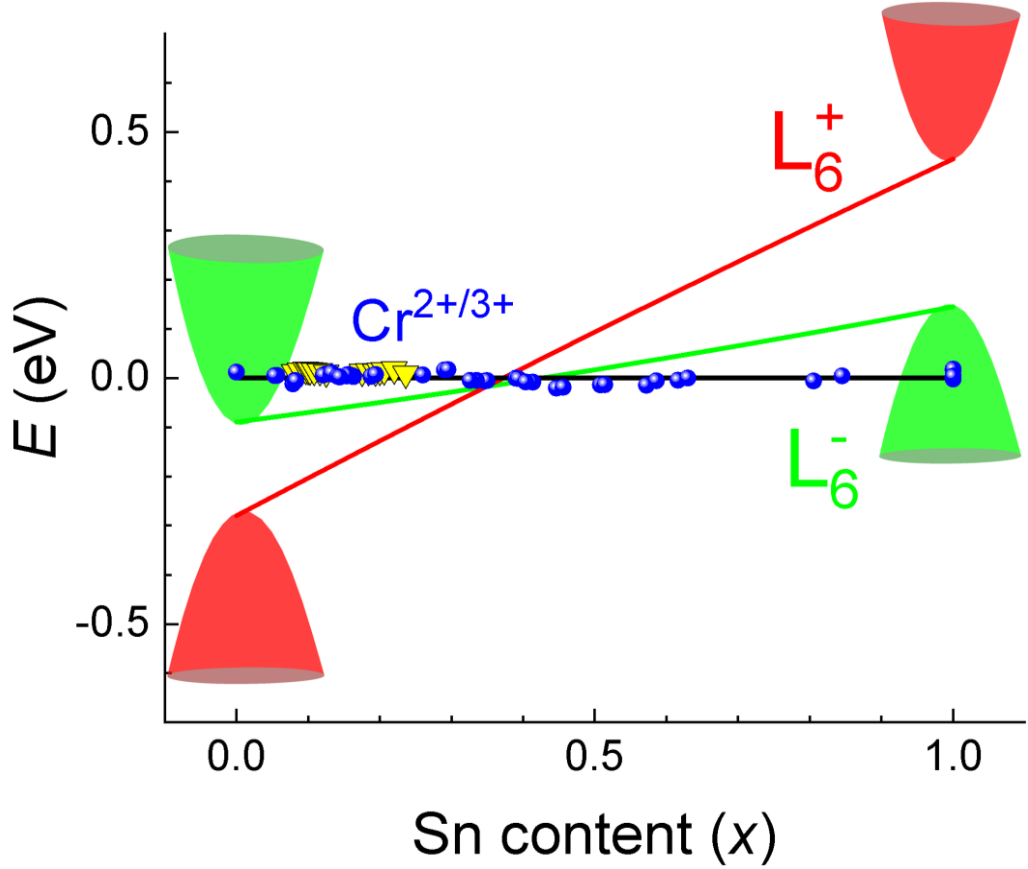

FIG. 12. Chromium $Cr^{2+/3+}$ level energy relative to the VBM and CBM of $Pb_{1-x}Sn_xTe$. Blue spheres are our experimental data calculated based on the carrier concentration at the Fermi level pinning conditions. For SnTe we place the $E_F$ calculated only for the two lowest hole densities depicted in Fig. 3. Yellow triangles are experimental results from Ref. [25]. The bands $L_6^+$ and $L_6^-$ are marked with red and green, respectively. Conduction and valence band lines are to scale, while the shapes of bands are symbolic. Values are given for $T = 4.2$ K. Zero energy is at the $Cr^{2+/3+}$ level.

The fitted energies of the $Cr^{2+/3+}$ level for in entire composition range of $Pb_{1-x}Sn_xTe$ under the pinning conditions are displayed in Fig. 12. Moreover, our transport measurements for PbSe presented in Fig. 2 allow us to establish the energy of the $Cr^{2+/3+}$ level in this host, and to include PbSe in the study of band offsets. We take the following energy gap values at liquid hellium temperatures: $E_{gap}(PbTe) = 0.19$ eV, $E_{gap}(SnTe) = 0.3$ eV and $E_{gap}(PbSe) = 0.15$ eV [59]. The resulting band-offsets at PbTe/SnTe, PbTe/PbSe and PbSe/SnTe interfaces are shown in Fig. 13 and given in Table III.

TABLE III. Conduction and valence band offsets in eV for PbTe/SnTe, PbTe/PbSe and PbSe/SnTe.

| heterostructure | $\Delta E(+)$ | $\Delta E(-)$ | VBO | CBO |
|---|---|---|---|---|
| SnTe/PbTe | 0.72 | 0.23 | 0.42 | 0.55 |
| PbSe/PbTe | 0.02 | - 0.02 | 0.02 | -0.02 |
| SnTe/PbSe | 0.70 | 0.25 | 0.4 | 0.57 |

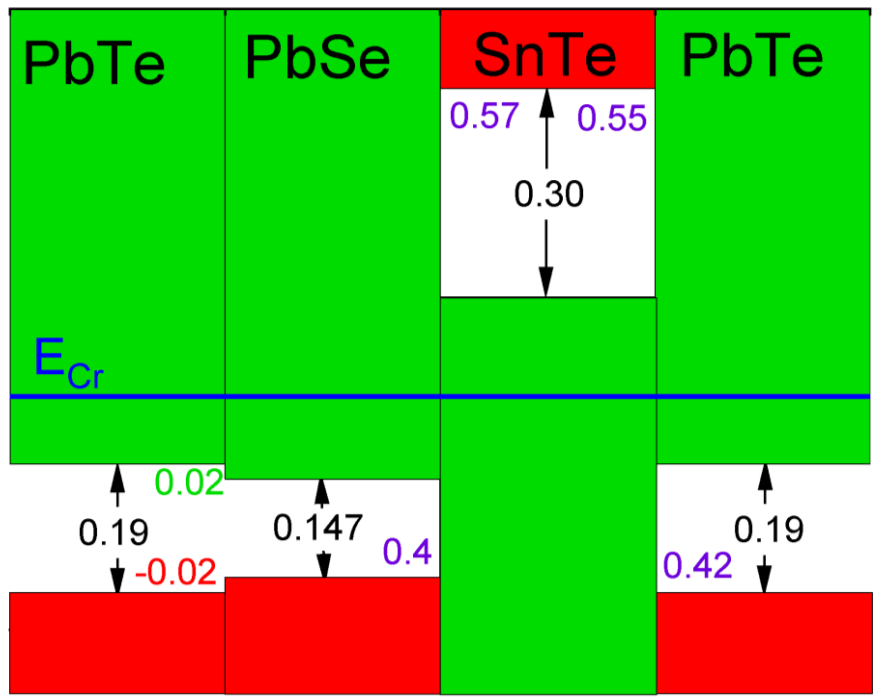

FIG. 13. Band offsets at PbTe/PbSe/SnTe interfaces. The $Cr^{2+/3+}$ level position is denoted with the solid blue line. All values are given in eV. Energy gap values are denoted by black arrows. The colors reflect the parity of the bands, as in Fig 12.

Comparing our results to the literature data, we observe that PbTe/(Sn,Pb)Te systems were investigated experimentally in several works, but the proposed values exhibit a relatively large spread. In Ref. [37] transport properties of (111) oriented PbTe/(Pb,Sn)Te superlattices with $x$=0.20 were investigated. Based on the experimentally determined In impurity level they proposed that the values extrapolated to the PbTe/SnTe interface are $\Delta E(+) = 0.7$ eV and $\Delta E(-) = 0.2$ eV (i.e., VBO=0.4 eV and CBO=0.5 eV). Similar values, $\Delta E(+) = 0.76$ eV and $\Delta E(-) = 0.27$ eV, (VBO=0.46 and CBO=0.57) were obtained in Ref. [38] from transport measurements. These values were next used in Ref. [36] and [60], providing an accurate description of their experimental results. Our band offsets are in good agreement with these data. Comparable results were also obtained in Ref. [56], which investigated magnetooptical properties of (111) PbTe/SnTe multiple quantum wells. These authors observed that the experimental data can be fitted with a comparable accuracy using two sets of the values, namely either $\Delta E(+) = 0.7$ eV and $\Delta E(-) = 0.17$ eV (VBO=0.4 eV and CBO=0.5 eV), which agrees with the values above, or with somewhat lower $\Delta E(+)$=0.4 eV and $\Delta E(-)$=-0.13 eV (VBO=0.06 eV and CBO=0.2 eV). In contrast, much higher offsets were recently determined by X-ray photoelectron spectroscopy in Ref. [40], $\Delta E(+) = 1.55$ eV and $\Delta E(-) = 1.05$ eV (i.e., VBO=1.37 eV and CBO=1.23 eV) for a (111)-oriented PbTe/SnTe heterojunction. Finally, we observe that a similar scheme to that of Fig. 12 was suggested based on the results obtained with other TM dopants [12].

Our band offsets for the PbTe/PbSe heterostructure are CBO=0.02 eV and VBO= -0.02 eV, respectively. These values are in good agreement with both the result of Ref. [61], CBO= 0.04 eV, and with the theoretical value CBO=0.1 eV of Ref. [59]. However, there is an uncertainty in literature about exact value of energy gap for PbSe [1]. For our purposes, if we take $E_{gap}$(PbSe) = 0.165 eV, the VBO at the PbTe/PbSe interface vanishes, while CBO remains unchanged, since it is determined by the value of the Cr level realtive to CBM.

# V. THEORY

We begin this Section by recapitulating the essential features of the band structures of the pure hosts, and then move to the electronic structures of the substitutional Cr ion in PbTe and SnTe, and next to these of the interstitial Cr in both hosts. Finally, we consider formation energies of substitutional Cr ions and of the cation vacancies in both PbTe and SnTe, which allows to estimate the equilibrium concentrations of all the considered defects.

## A. Band structure of PbTe and SnTe

PbTe is a direct band gap semiconductor with the band gap at the $L$ point of the BZ. The wave functions of the VBM are of the $L_6^+$ symmetry, i.e., they are even with respect to inversion with the center at the cation site, and are mainly composed of the $s$(Pb) and $p$(Te) orbitals. The CBM wave functions are of the $L_6^-$ symmetry, odd with respect to inversion, and they are mainly composed of $p$(Pb) and $s$(Te) orbitals. Because of the inversion symmetry of the rock-salt structure, the bands of both PbTe and SnTe are spin degenerate, and both energy and orbital composition of spin partners are identical. Four $L$ points of the face centered cubic BZ are folded to a point denoted by "$L$" in the supercell BZ. After the folding, both the VBM and CBM are 4-fold degenerate at "$L$" neglecting spin (i.e., 8-fold degenerate with spin). In SnTe with the so-called inverted band structure, the order of the lowest conduction and the highest valence band is reversed, and both the VBM and CBM are slightly shifted away from the $L$ point, in agreement with Ref. [1]. The calculated band gaps are $E_{gap}$(PbTe) = 0.2 eV and $E_{gap}$(SnTe) = 0.3 eV, what agrees well with experiment [1,62].

Figure 14 presents the electronic configuration of the $d$ shell of Cr against the band structure of PbTe and SnTe. In Fig. 15 (a) we show the energy bands for PbTe supercells along the [111] direction of the BZ.

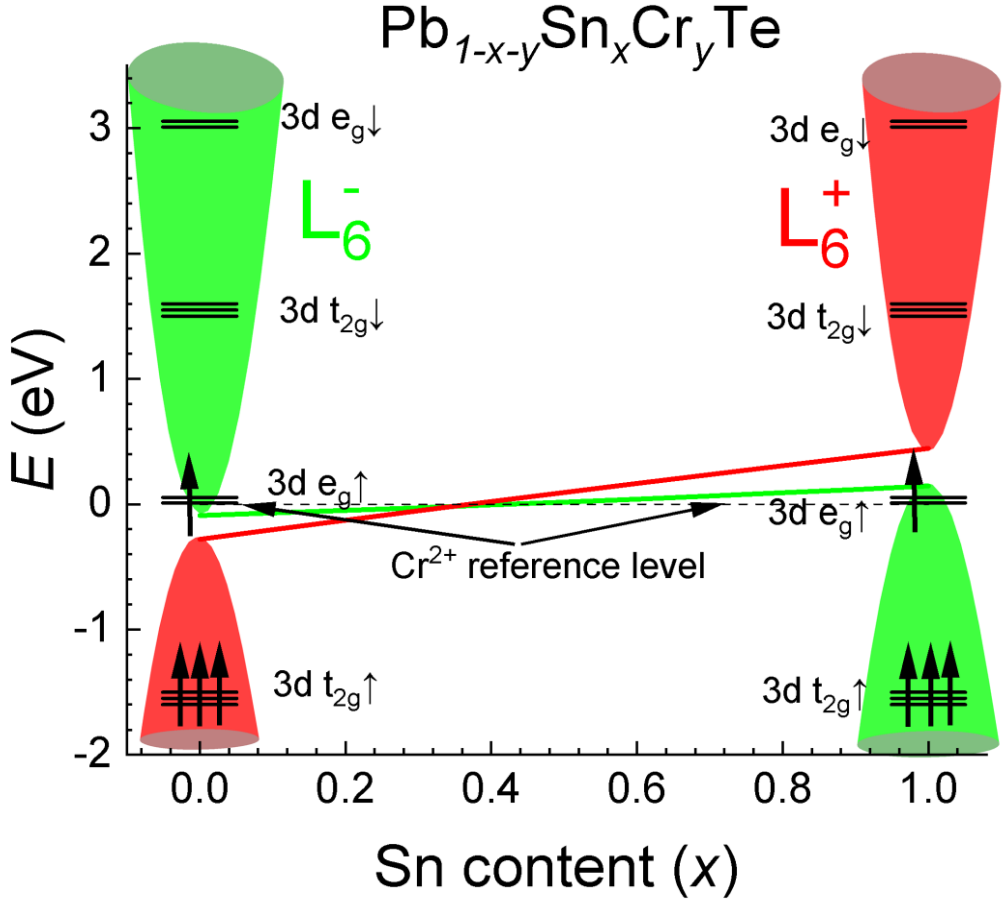

Fig. 14 Electronic configuration of octahedral crystal field and spin-exchage split Cr *3d* electron shell superimposed onto PbTe and SnTe band structure.

**B. Electronic structure of Pb$_{1-y}$Cr$_y$Te**

The electronic configuration of an isolated Cr atom is $3d^54s^1$. In a rock salt host, the *d*-shell of the substitutional Cr is split by the octahedral crystal field into $t_{2g}$(Cr) triplet, and $e_g$(Cr) doublet higher in energy. The crystal field splitting amounts to about 1.5 eV. Both multiplets are spin split by about 3 eV by the exchange coupling. Depending on the charge state, the Cr-derived spin-up bands are either partially or fully occupied, whereas the spin-down bands are degenerate with conduction bands and remain empty. As a consequence, the finite spin polarization of Cr leads to its high spin state. An analogous situation occurs for Cr in SnTe.

Figure 15 (b-d) shows the $e_g$(Cr)-derived band (referred to as the $e_g$ band in the following) for three charge states of Cr, together with the host energy bands of PbTe. The Cr-derived spin-up triplet is situated about 1.5 eV below the VBM, and is not visible in Fig. 15. The $e_g$ spin-up band overlaps with the bottom of the conduction band, and its center is about 50 meV above the CBM. This demonstrates that Cr is a resonant donor. Thanks to the compactness of the atomic 3$d$(Cr) orbitals and a weak hybridization with the conduction bands, this band is almost dispersionless. Cr is in the neutral charge state in the $d^4$ configuration, when $e_g$ is occupied by one electron. By comparing Figs. 15 (b), (c) and (d) we see that the energy of $e_g$ depends on its occupation, which will be addressed below.

A short comment regarding the convention used to label energy bands: The band denoted as "0" is *formally* the highest occupied one, and it is depicted by red dashed lines in Fig. 15 and below. Thus, in the case of a perfect host, the band "0" is the topmost valence band, Fig. 15 (a). A defect can induce a level in the band gap, and if it is a singlet occupied by one electron, it would be labeled "0" according to our convention. The case of Cr in PbTe, a resonant donor, is more complex, since it induces bands which overlap with the host bands, and therefore electrons are partially filling all the relevant bands up to the Fermi energy $E_F$ according to the Fermi-Dirac distribution. The actual concentration of conduction electrons in our model is determined by the size of the supercell used in the calculations and the energy of the resonant donor relative to the CBM; in our case of the 216 atom supercell, the Cr composition is 1/108, what corresponds to the dopant concentration of $1.5\times10^{20}$ cm$^{-3}$. Since the $e_g$ band is about 0.05 eV above the CBM, only a small fraction of Cr ions are ionized, and most of them are in the Cr$^{2+}$ state.

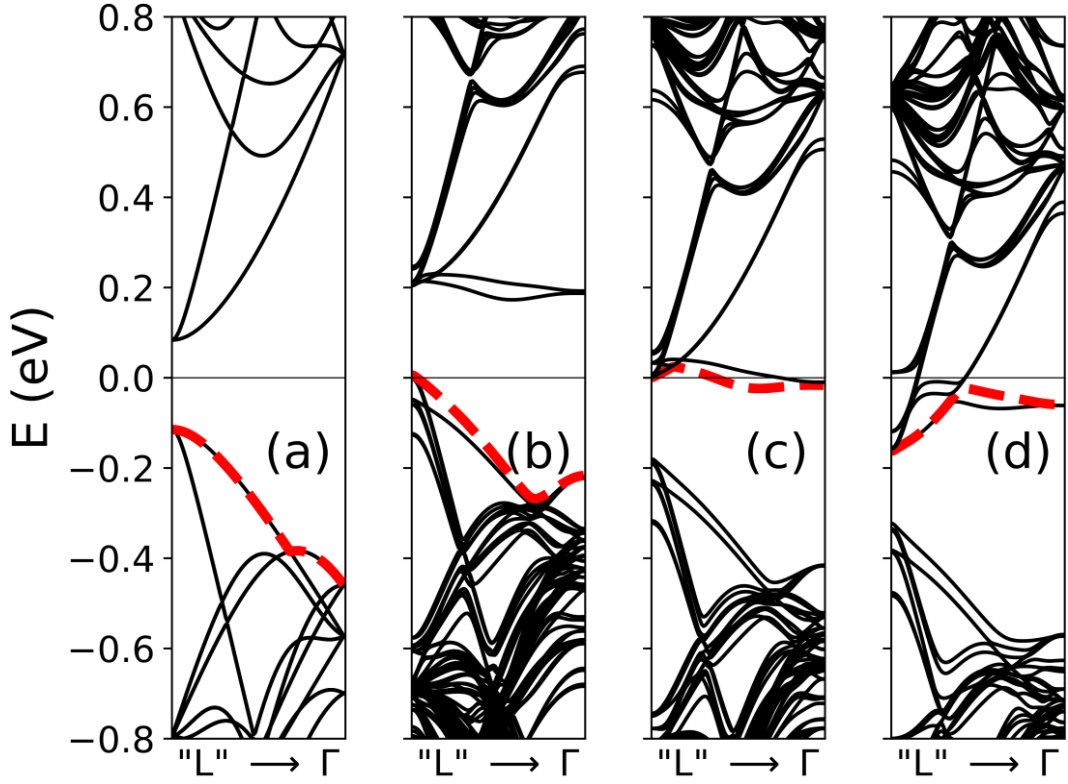

FIG. 15 Energy bands of (a) PbTe, and of $Pb_{1-y}Cr_yTe$ for three charge states of chromium: (b) $Cr^{3+}$, (c) $Cr^{2+}$, and (d) $Cr^{1+}$. Zero energy corresponds to the Fermi energy. The red dashed lines denote the highest occupied bands, see text.

The resonant character of $e_g$ is clearly reflected in the orbital composition of the bands close to $E_{gap}$, presented in Fig. 16 for the case of $Cr^{2+}$ shown in Fig. 15 (c). The orbital composition is calculated according to Ref. [63]. In the analysis, we consider the 8 lowest conduction bands (2-9) with the index increasing with energy, while the bands from -1 to -8 are the highest valence bands with the index decreasing with energy. The bands 0 and 1 are the Cr-derived $e_g$ bands, and the band 0 is the highest occupied one.

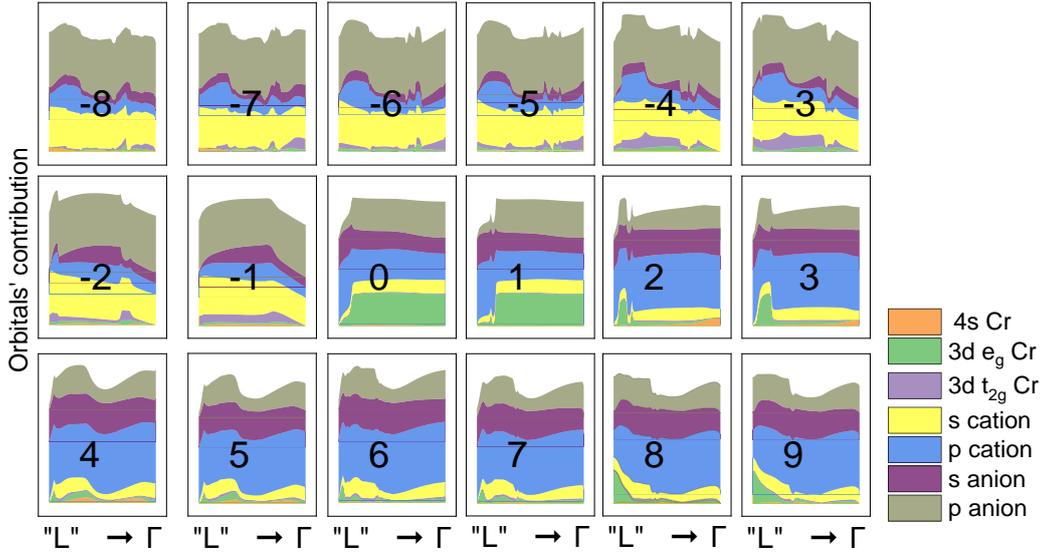

FIG. 16 Contributions of atomic orbitals to the wave functions in dependence on $k$ vector along [111] direction for $Pb_{1-y}Cr_yTe$. The bands from -8 to -1 are the valence bands, the bands 0 and 1 are the $e_g$ $d$(Cr)-derived, and from 2 to 9 are conduction bands.

In Fig. 16 the green color indicates the contribution of the $e_g$ $d$(Cr)-derived atomic states, which almost dominates both the band 0 and its upper partner, the first empty band 1. In the close vicinity of the "$L$" point, the lowest host conduction bands and $e_g$ cross, which is displayed by the abrupt change of the orbital content of the subsequent energy bands. The crossing is manifested as a sharply peaked contribution of $d$(Cr) to the bands 2 and 3 which vanishes at "$L$", and the high contribution of $d$(Cr) visible for the bands 8 and 9 situated about 50 meV above the CBM. These results confirm that Cr is a resonant donor in PbTe.

We now turn to the charge state of the Cr impurity. This issue is more complex than that of a standard donor with a level in the band gap. Because of the Cr resonance character, an isolated Cr autoionizes giving an electron to the bottom of the conduction band, and it assumes the $Cr^{3+}$ charge state. With the increasing Cr concentration, the Fermi level increases, reaching the resonance level for a critical concentration when electrons fill the bottom of the conduction band up to the $e_g$ band, and some Cr ions assume the $Cr^{2+}$ charge state. Consequently, the concentration of conduction electrons is determined by the energy of $e_g$ relative to the CBM. For high Cr concentrations the Fermi level pinning takes place, and two charge states of Cr dopants, $Cr^{3+}$ and $Cr^{2+}$, coexist. The considered case of a Pb$_{107}$Cr$_1$Te$_{108}$ supercell corresponds to the Cr concentration of about $1.5 \times 10^{20}$ cm$^{-3}$, i.e., to a relatively high doping level. As it follows from Fig. 15 (c), in this case the lower partner of the $e_g$ band is almost completely filled, most of the Cr ions are in the Cr$^{2+}$ (i.e., $d^4$) state, and indeed the calculated spin S=2.

In real crystals, additional acceptors or donors can be present, which can change $E_F$ and the charge state of Cr dopants. We model such a situation by subtracting (or adding in the case of a donor) one electron from the supercell with a Cr ion. In other words, we assume the presence of a "generic" acceptor or donor in the supercell. Figure 15 (b) presents energy bands in the case when one electron is subtracted from the supercell. As a result, the $e_g$ doublet is empty, and Cr is in the Cr$^{3+}$ state with spin $S=3/2$. Since PbTe typically is $p$-type because of the presence of electrically active cation vacancies, some amount of Cr$^{3+}$ is expected to always occur (provided that there are no donors in the specimens).

Figure 15 (d) presents the band structure when one electron is added to the supercell, and thus both partners of the $e_g$ band are almost totally occupied with two electrons. In this case, the resonant character of Cr and its autoionization are more clear than in Fig. 15 (c) because the energy of $e_g$ is about 150 meV above the CBM. This implies co-existence of Cr$^{2+}$ and Cr$^{1+}$ ions. As it follows from Fig. 15, the actual energy of the $e_g$ doublet relative to the CBM depends on its occupation, i.e., on the Cr charge state. The effect is dictated by the intra-center Coulomb repulsion between the $d$(Cr) electrons. In the case of Cr$^{1+}$, the energy of the fully occupied $e_g$ is higher than that of Cr$^{2+}$ by ~0.1 eV, which reflects the magnitude of the effective intrashell Coulomb repulsion within the $e_g$ band.

### C. Electronic structure of Sn$_{1-y}$Cr$_y$Te

Turning to Cr in SnTe, Fig. 17, we see that the energies of the $e_g$ bands are lower compared to those in PbTe. In particular, from Fig. 17 (c) it follows that in a perfect SnTe (i.e., without free carriers), Cr is a shallow resonant acceptor, with the $e_g$ band practically degenerate with the VBM and occupied with one electron. This demonstrates that Cr is in the Cr$^{2+}$ charge state with the corresponding $d^4$ configuration, as in PbTe. Accordingly, the spin of Cr is $S=2$. The resonant character of Cr in SnTe is confirmed by the decomposition of the wave functions displayed in Fig. 18. Indeed, close to the "$L$" point the Cr-dominated bands (0,1) cross the first two valence bands -1 and -2. We also see a weak hybridization of $d$(Cr) with the bands from -3 to -6. The dependence of the $e_g$ energy on the Cr charge state is similar to that of Cr in PbTe: adding one electron up-shifts the $e_g$ band by about 0.2 eV, while adding one hole to the system causes only a small down-shift of $e_g$.

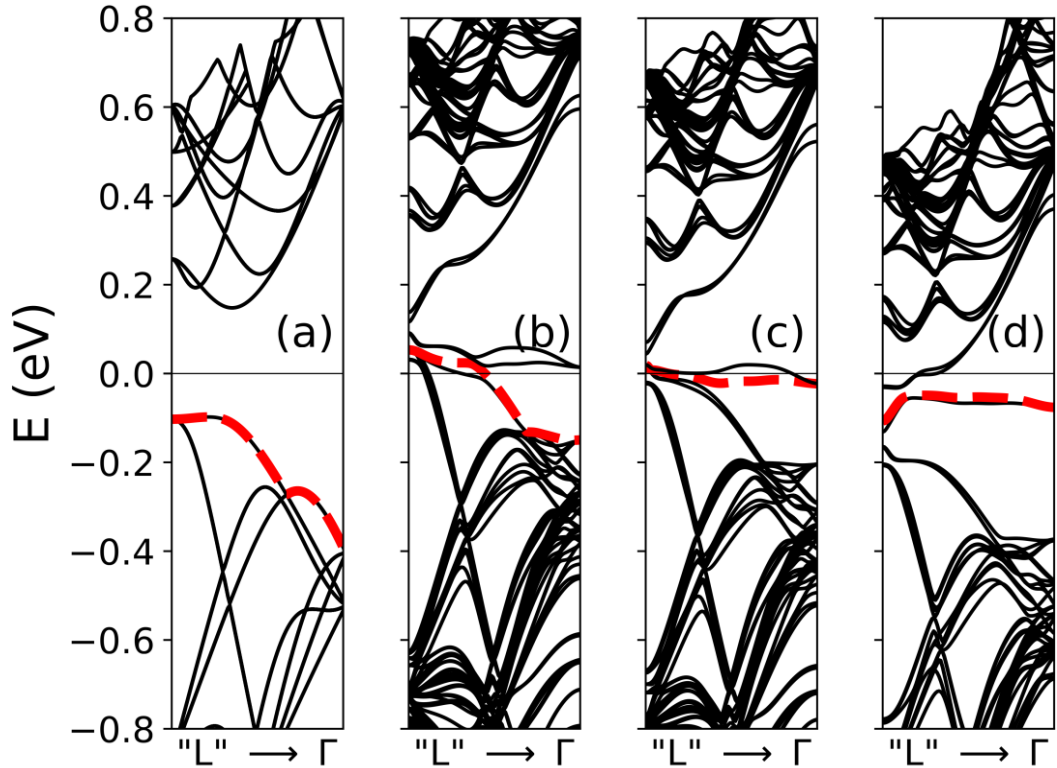

FIG. 17 Band structure of (a) SnTe, and $Sn_{1-y}Cr_yTe$ for three charge states of chromium, (b) $Cr^{3+}$, (c) $Cr^{2+}$, and (d) $Cr^{1+}$. Zero energy corresponds to the Fermi energy. Red dashed lines denote the last occupied bands, see text.

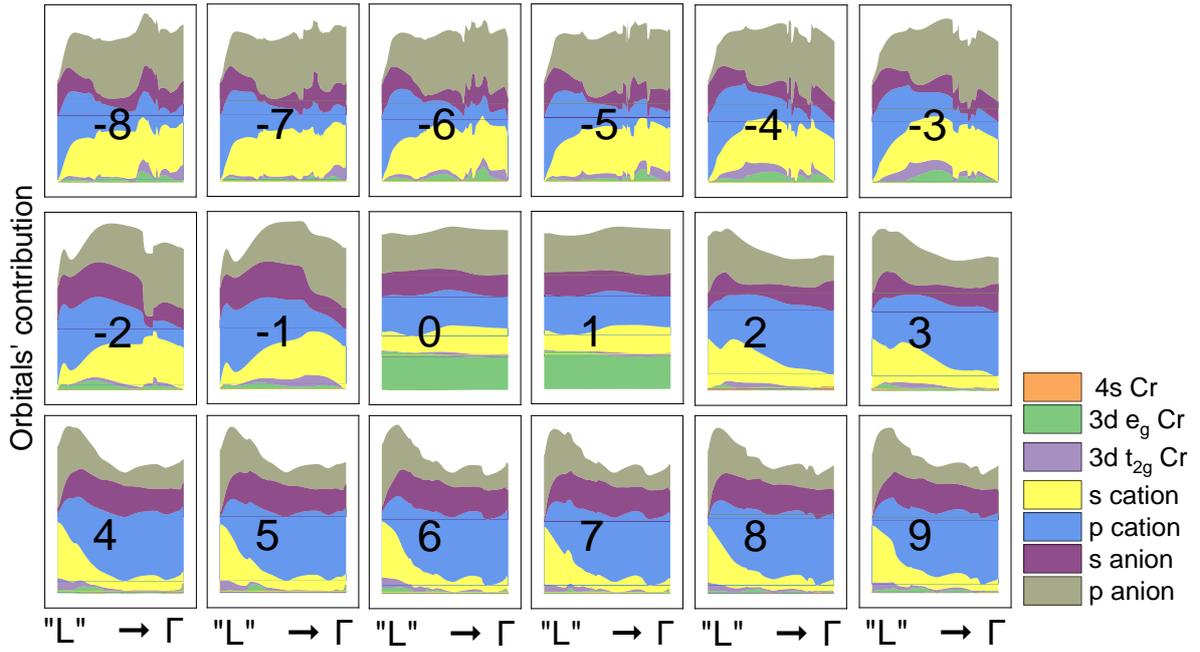

FIG. 18 Contributions of atomic orbitals to the wave functions as a function of the *k* vector along the [111] direction for $Sn_{1-y}Cr_yTe$. The bands from -8 to -1 are the valence bands, the bands 0 and 1 are $e_g$ *d*(Cr)-derived, and from 2 to 9 are conduction bands.

As it was extensively discussed in Ref. [63], the hybridization between a dopant and the host neighbors is determined by symmetries of the involved states. In the case of Cr, both *s*(Cr) and *d*(Cr) orbitals are even with respect to inversion, and therefore they couple to even parity wave functions $L_6^+$. Accordingly, the symmetry based arguments explain the large, comparable to the band gaps, values of both the energy shifts of the $L_6^+$ band extrema and their splittings by ~0.2 eV, reported in Figs. 15 and 17.

Finally, because the atomic radius of Cr is smaller than these of Pb and Sn, it is important to find equilibrium atomic configurations in the vicinity of Cr. In both hosts the relaxation consists in reducing the Cr-Te distance, what affects the energy of the $e_g$ (Cr) level. In PbTe, the equilibrium Cr-Te bond length, 2.88 Å, is shorter than the Pb-Te bond, 3.23 Å, by about 11 %, and the corresponding energy shift of $e_g$ is 0.3 eV relative to host bands. Indeed, without the configuration optimization, Cr is a resonant acceptor, degenerate with the VBM. A smaller impact of relaxation is obtained for Cr in SnTe, where the radius misfit between Cr and Sn is lower, and the bond is reduced by 9 % from the ideal value 3.15 to 2.86 Å. The relaxation energy (energy gain) is 1 eV for $Pb_{1-y}Cr_yTe$ and 0.8 eV for $Sn_{1-y}Cr_yTe$.

## D. Interstitial Cr in PbTe and SnTe

A second possible location of Cr ions is the interstitial position. In the rock salt structure, there is one type of such sites, shown in Fig. 19. In the non-relaxed configuration it is a cube with 4 cations and 4 anions in the corners, and the tetrahedral symmetry relative to the cube center. The interstitial is therefore 8-fold coordinated. In such a configuration, the $d$-shell is split by the crystal field into a doublet, and a higher in energy triplet, i.e., the order of levels is reversed compared to the case of a substitutional Cr of octahedral symmetry. After atomic relaxation, the local tetrahedral symmetry of Cr is reduced to $C_{3v}$ because of the actual configuration of cations and anions at the cube corners. The equilibrium bond lengths $b$ are given in Table IV. As expected, the $b$(Cr-Te) bonds are close in both hosts, while $b$(Cr-cation) are not. As it follows from the Table IV, the displacements of Cr neighbors lowers the point symmetry from $T_d$ to $C_{3v}$. An analogous symmetry breaking is found in the case of $Pb_I$ in PbTe [64].

Table IV. The equilibrium bond lengths between $Cr_I$ and its neighbors. The $b_1$ bonds are along the [111] direction, and $b_2$ denote the three remaining tetrahedrally oriented bonds. All lengths are in Angstroms.

|      | $b$ ideal | $b_1$(Cr-Te) | $b_2$(Cr-Te) | $b_1$(Cr-cation) | $b_2$(Cr-cation) |
|------|-----------|--------------|--------------|------------------|------------------|
| PbTe | 2.80      | 2.74         | 3 × 2.83     | 2.92             | 3 × 3.01         |
| SnTe | 2.73      | 2.75         | 3 × 2.84     | 2.81             | 3 × 2.86         |

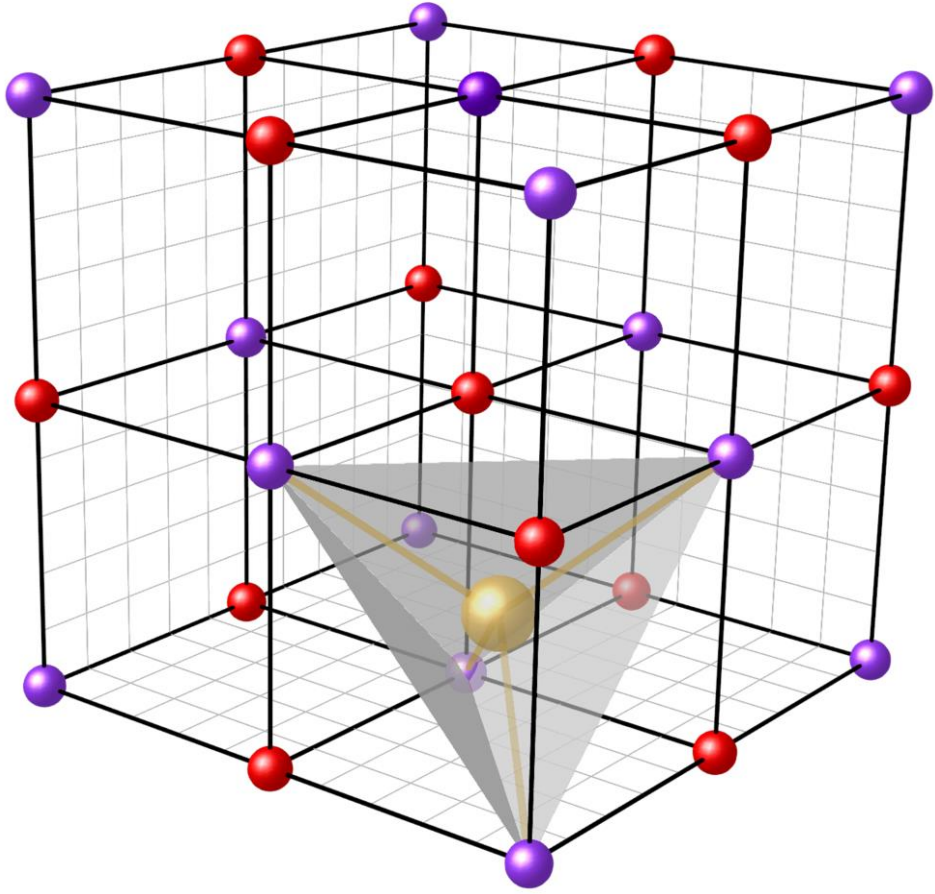

FIG. 19. PbTe unit cell with a Cr atom (marked in gold) in the tetrahedral interstitial position with four bonds along the [111] direction.

The calculated energy bands of a SC with an interstitial Cr, $Cr_I$ in PbTe, together with the corresponding densities of states, are shown in Fig. 20. As it follows from the DOS projected onto the $t_{2g}$ and $e_g$ Cr states, in PbTe the spin up $e_g$ doublet and the spin up $t_{2g}$ triplet are degenerate with the valence bands and fully occupied, while the spin-down $d(Cr)$-derived bands are located above the CBM and empty. Consequently, the $Cr_I$ is in the $Cr^{1+}$ ($d^5$) configuration. Moreover, there is one electron in the conduction band of our supercells, indicative that $Cr_I$ is a single donor. The PDOSs shown in Fig. 20 reveal a substantial hybridization of $d(Cr_I)$ states with the host bands, which is reflected by the widths of the $d(Cr)$-derived peaks amounting to 1-2 eV. Because of this pronounced hybridization, it is not possible to reasonably estimate neither the corresponding crystal field splitting nor the spin splitting. We only note that the sharp peak in PDOS of $t_{2g}$ at 0.3 eV below the VBM is visible as a flat band in Fig. 20 (a).

Electronic structure of $Cr_I$ in SnTe presented in Fig. 21 is quite close to that of $Cr_I$ in PbTe. Both the spin up $e_g$ doublet and the $t_{2g}$ triplet are degenerate with the valence bands, the hybridization between $d(Cr)$ and the host states is as pronounced as in PbTe, the spin down $d(Cr)$ states are above the CBM, and $Cr_I$ is a single donor in the $Cr^{1+}$ ($d^5$) electronic configuration.

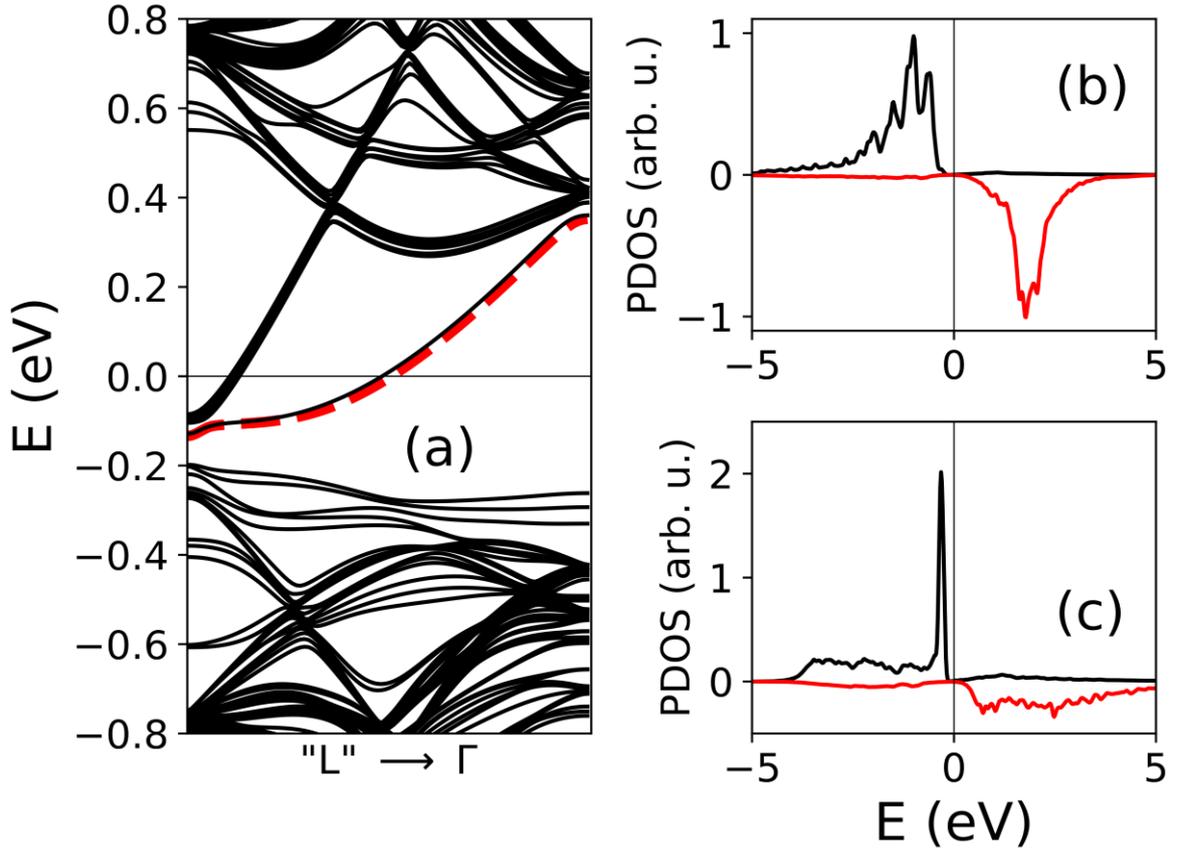

FIG. 20 (a) Energy bands of PbTe with intersititial Cr. Red dashed line denotes the highest occupied band. The DOSs projected onto the $e_g$ and $t_{2g}$ Cr orbitals are shown in (b) and (c), respectively, and the positive and negative values describe the spin up and spin down states, respectively. Zero energy corresponds to $E_F$.

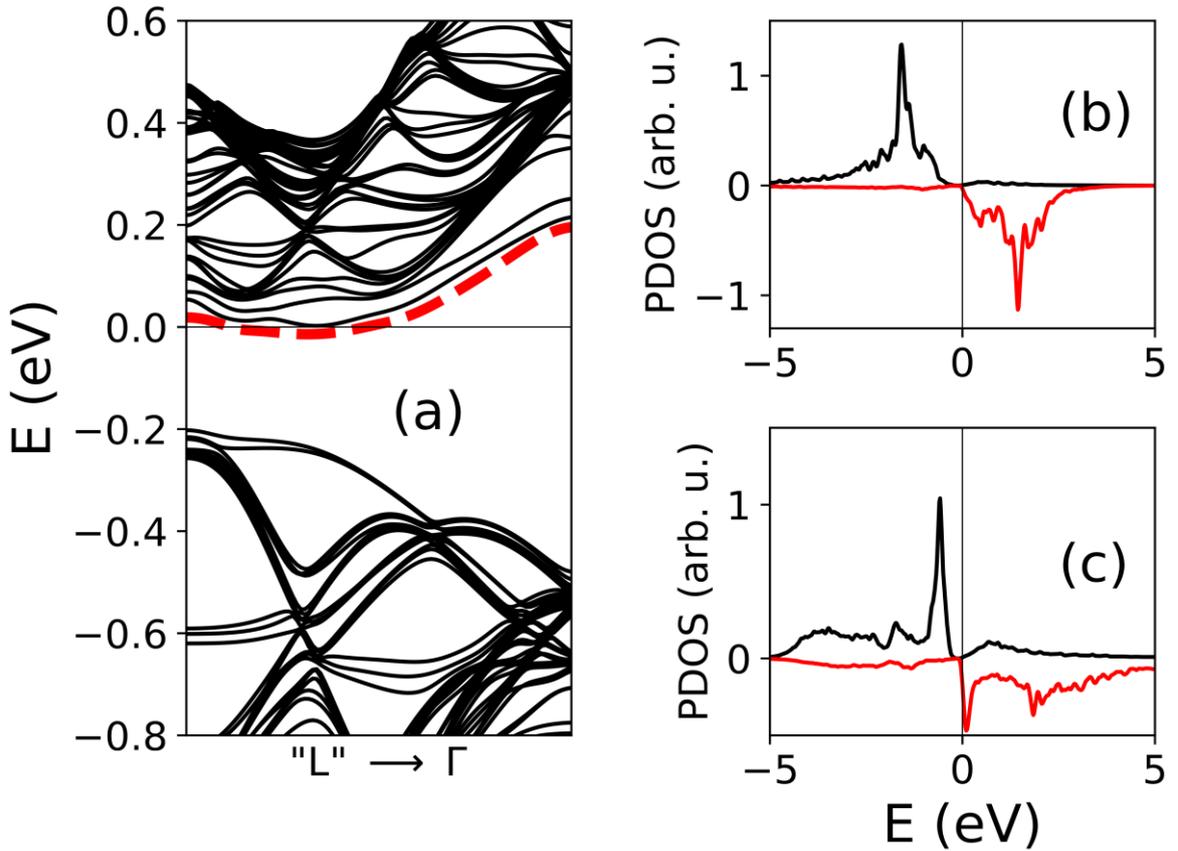

FIG. 21 (a) Energy bands of SnTe with interstitial Cr. Red dashed line denotes the highest occupied band. The DOSs projected onto the $e_g$ and $t_{2g}$ Cr orbitals are shown in (b) and (c), respectively, and the positive and negative values describe the spin up and spin down states. Zero energy corresponds to $E_F$.

### E. Cation vacancies

As is was mentioned above PbTe and SnTe, exhibit a tendency to be *p*-type due to the presence of cation vacancies $V_{cation}$. This is true in particular in the case of SnTe, where the hole concentrations *p* of the order $10^{20} – 10^{21}$ cm$^{-3}$ are a rule, while in PbTe a typical hole concentration is about three orders of magnitude lower. Theoretical studies of point native defects in PbTe and SnTe performed in Refs [64–71] explain these findings. Indeed, the calculated formation energies of cation vacancies are low, and they are double acceptors. Moreover, in most growth conditions formation energies of the remaining native defects are higher, and these defects are donors. Finally, formation energies of $V_{Pb}$ in PbTe are higher than those of $V_{Sn}$ in SnTe, again in accord with experiment. Based on these findings we limit our analysis to the cation vacancies only. Since both substitutional and interstitial Cr ions are donors, the cation vacancies can play a role of compensating native defects, which can additionally increase the solubility of Cr in both hosts. Energy bands of both SnTe and PbTe with a cation vacancy are displayed in Fig. 22. There are no vacancy-induced levels in the band gap, and therefore in both compounds the cation vacancies act as resonant double acceptors which fix the Fermi level below the VBM. cation vacancies act as resonant double acceptors which fix the Fermi level below the VBM.

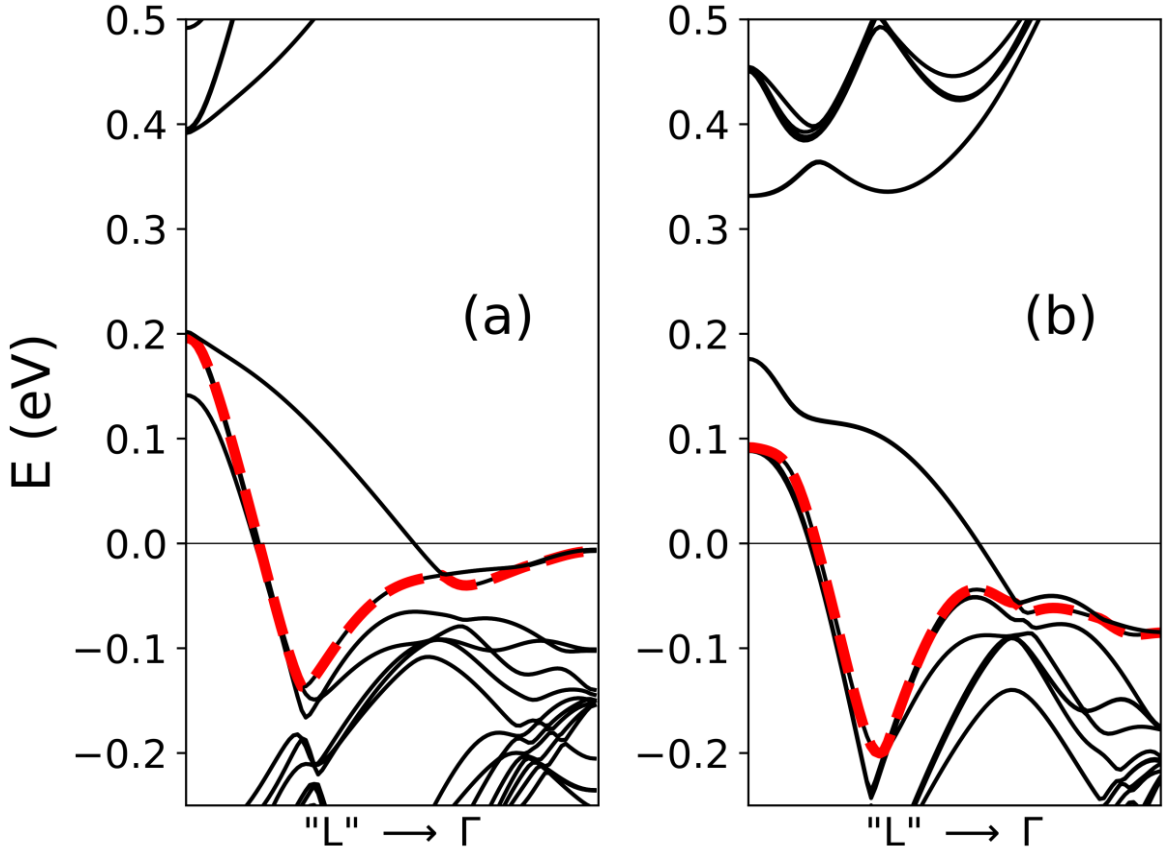

FIG. 22. Calculated energy bands of (a) PbTe and (b) SnTe supercells with a cation vacancy. Zero energy is at the Fermi level. Red dashed lines denote the highest occupied bands, see text.

In our samples the Cr ions and $V_{cat}$ coexist, and both defects are present at high concentrations. The actual charge states of Cr and $V_{cat}$ are determined by the relative energies of their levels. While in PbTe the situation is clear since $Cr_{Pb}$ is a donor and $V_{Pb}$ is an acceptor, in SnTe and Sn-rich PbSnTe samples the situation can be ambiguous, because both $V_{Sn}$ and $Cr_{Sn}$ introduce levels degenerate with the valence bands. To assess the charge state of Cr ions in our samples, let us examine a SC simultaneously containing both centers. The partial densities of states corresponding to $e_g(Cr)$ and $t_{2g}(Cr)$ states of both $Cr_{Sn}$ and $Cr_I$ are shown in Fig. 23. The calculated PDOSs correspond to $Cr_{Sn}^{3+}$ and $Cr_I^{1+}$, respectively, what follows from the position of $E_F$ relative the Cr-induced states. This shows that $Cr_{Sn}$ acts as a donor compensating holes.

The theoretical results, and in particular these shown in Fig. 23, are vital for the interpretation of the experimental data. Indeed, experiment demonstrates a simultaneous presence of $Cr^{1+}$, $Cr^{2+}$ and $Cr^{3+}$ together with free carriers (Fig. 10) in the whole composition range of $Pb_{1-x}Sn_xTe$. Our calculations allow for the unequivocal distinction between the contributions provided by Cr ions in the substitutional and in the interstitial positions. This is because the only expected charge state of the interstitial Cr is $Cr^{1+}$, while the $Cr^{3+}$ and $Cr^{2+}$ charge states correspond to the substitutional Cr.

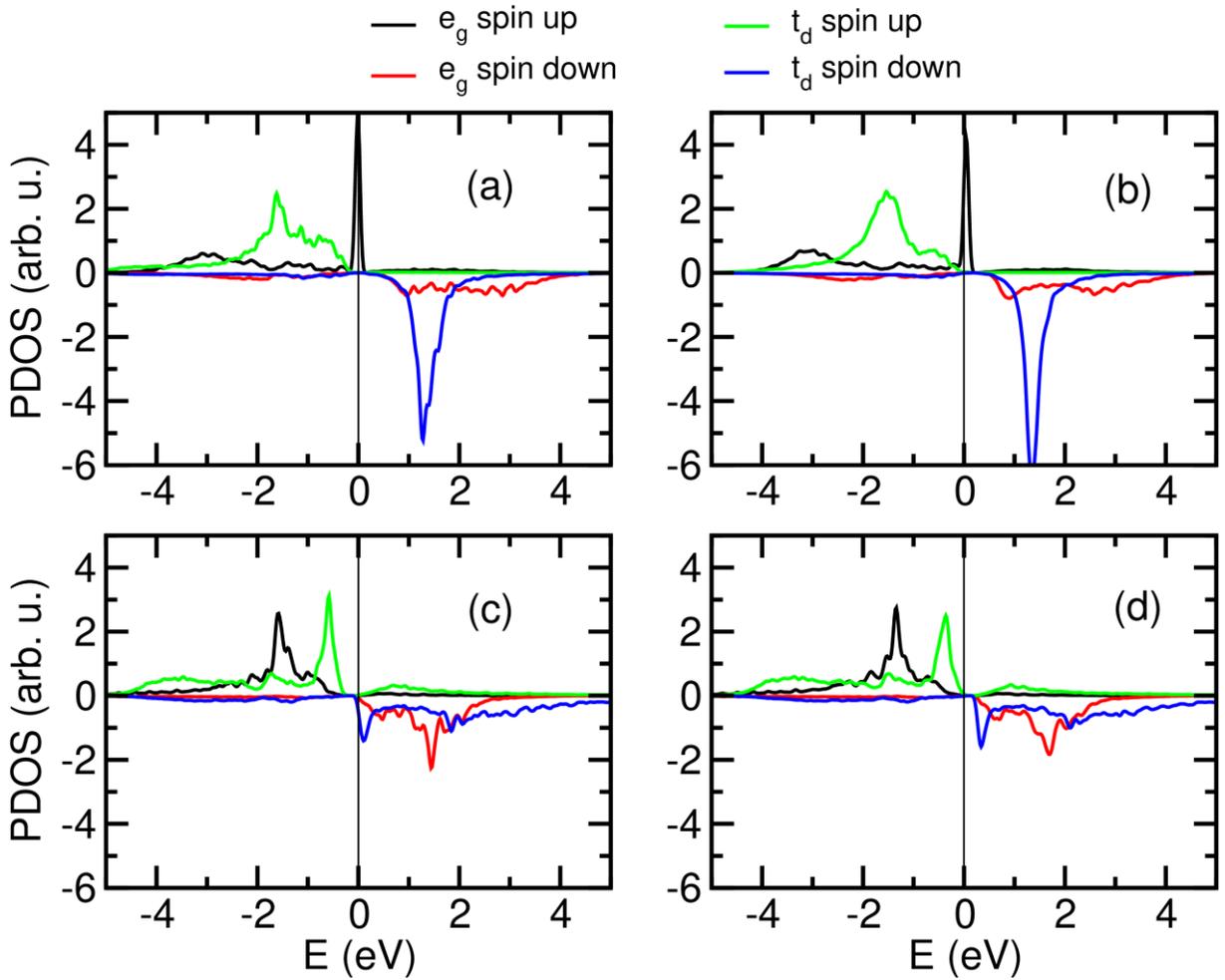

FIG. 23 Partial density of states of d(Cr) orbitals in the case of (a) $Cr_{Sn}$ and (b) $Cr_{Sn}$ and $V_{Sn}$ in the supercell, and (c) $Cr_I$ and (d) $Cr_I$ and $V_{Sn}$ in the supercell. Zero energy is at the Fermi level. Positive and negative values correspond to spin-up and spin-down states.

## F. Formation energies of Cr dopants and cation vacancies

Formation energy $E_{form}(D^q)$ of a defect D in the charge state $q$ (relative to isoelectronic charge state) is calculated according to [72,73]:

$$E_{form}(D^q) = E(D^q) - E_{ideal} + \Delta N_i m_i + qE_F \qquad (3)$$

Here, $E_{ideal}$ and $E(D^q)$ are the total energies of a supercell without and with a defect, $\Delta N_i$ is the number of type $i$ atoms added to ($\Delta N_i < 0$) or removed from ($\Delta N_i > 0$) the supercell during the formation of the defect, and $E_F$ is the Fermi energy relative to the energy of the VBM. $q$ is the defect charge state, and $q = 0$ for $Cr^{2+}$. $\mu_i$s are the variable chemical potentials of atoms in the solid, which in general are different from $\mu_i$(bulk) of the assumed standard state of elements, i.e., the diamond phase of Sn, the face centered phase of the cubic Pb, and the body centered cubic phase of Cr. Moreover, the chemical potentials satisfy the relation $\mu_{Sn} + \mu_{Te} = \Delta H_f(SnTe) + \mu_{Sn}(bulk) + \mu_{Te}(bulk)$, where $\Delta H_f(SnTe)$ is the heat of formation of SnTe, and an analogous formula holds for PbTe. Formation energy includes also the potential alignment and the image charge correction energy, see [72] [74]. Their form and value are analyzed in detail in Ref. [72] [73]. Due to the relatively large supercell used here and the very large dielectric constant of $Pb_{1-x}Sn_xTe$, the value of both corrections are about 0.01 eV, which is neglected here.

The calculated formation energies of vacancies are $E_{form}(V_{Sn})=0.05$ eV, and $E_{form}(V_{Pb})=1.55$ eV. Both values correspond to the cation-rich conditions, the -2 charge states of defects, and for the $E_{Fermi}$ located at the VBM. They are in reasonable agreement with the values reported in Refs [64–71].

Table IV. Formation energies in eV of the substitutional $Cr_{cation}$, the interstitial $Cr_I$, and the cation vacancies in PbTe and SnTe.

|  | PbTe | SnTe |
|---|---|---|
| $Cr_{cation}$ | 1.9 | 1.1 |
| $Cr_I$ | 1.1 | 1.15 |
| $V_{cation}$ | 1.55 | 0.05 |

As it follows from Table IV, in SnTe the interstitial and the substitutional incorporations are almost energetically equivalent, and in PbTe the interstitial Cr is energetically preferred. Considering the cation vacancies, we find that their formation energies are very low in SnTe, and much higher in PbTe, which agrees quantitatively with the experimental observations.

## V. CONCLUSIONS

In this paper we present the results of experimental (electron transport, magnetization, EPR) and theoretical (DFT) studies of $Cr^{2+/3+}$ resonant dopant states in $Pb_{1-x}Sn_xTe$ semiconductors across a full range of Sn content. It covers the band inversion region with a disorder induced Weyl semimetal phase as well as both topologically trivial (PbTe-rich) and topological crystalline insulator (SnTe-rich) regions. The experimentally observed $n$-type doping and Fermi level pinning in $Pb_{1-x}Sn_xTe$ doped with chromium is discussed in a realistic model involving electrically active $Cr^{3+}$ donors, neutral $Cr^{2+}$, as well as $p$-type background doping by metal vacancies. The determined Fermi energies are strongly Cr concentration dependent what enables to trace the location of Cr level in the whole range of Sn composition in $Pb_{1-x}Sn_xTe$ crystals. It provides possibility to control carrier concentration in both $p$- and $n$-type materials, which are very important in applications related to thermoelectricity. In the present paper we also develop one consistent procedure for magnetization data analysis in order to determine spin and charge states of chromium dopant in IV-VI compounds. This way we extend the existing method recently applied by some of us exclusively for $Pb_{1-y}Cr_yTe$. We propose this approach as a unique tool for determination of the exact concentration of paramagnetic Cr dopants in particular charge/spin states of ($Cr^{3+}$, $Cr^{2+}$, $Cr^{1+}$), thus avoiding difficulties of other detection methods which always account also for Cr-Te precipitates. We emphasize that this approach may be especially useful for analyzing compounds with low solubility limit of magnetic dopants. Using this method we also determine the concentration of $Cr^{2+}$ ions, whose presence evades detection by conventional EPR. Our *ab-initio* calculations confirm that $Cr^{3+}$ and $Cr^{2+}$ charge states correspond to substitutional Cr. Moreover, they also point out that $Cr^{1+}$ is the equilibrium charge state of interstitial donors. Formation energies of the substitutional and interstitial Cr are close, in agreement with their similar experimental concentrations. This interstitial incorporation of Cr in IV-VI compounds was previously not considered. Theoretical studies also give us opportunity for recognition of the role of parity of conduction/valence bands and *3d* states of Cr in hybridization of transition metal and band states, thus reshaping the near-gap states. Finally, the fact that transition metal impurities may serve as universal reference levels in various families of compounds enables us to determine the band-offsets in the PbTe/SnTe, PbTe/PbSe, and PbSe/SnTe heterojunctions. This opens the new field for developing devices based on IV-VI compounds including topological crystalline insulators and 2D heterostructures.

## ACKNOWLEDGEMENTS

This study has been supported by the National Science Centre (Poland) through project OPUS (UMO – 2017/27/B/ST3/02470) and by the National Science Centre for Development (Poland) through grant TERMOD No TECHMATSTRATEG2/408569/5/NCBR/2019 and by the Foundation for Polish Science project "MagTop" no. FENG.02.01-IP.05-0028/23 co-financed by the European Union from the funds of Priority 2 of the European Funds for a Smart Economy Program 2021–2027 (FENG).

# SUPPLEMENTAL MATERIAL
## for
## Cr resonant impurity for studies of band inversion and band offsets in IV-VI semiconductors

A. Królicka [1]*, K. Gas [1,2,3], W. Dobrowolski [1], H. Przybylińska[1],
Y. K. Edathumkandy[1], J. Korczak [1,4], E. Łusakowska [1], R. Minikayev [1],
A. Reszka[1], R. Jakieła [1], L. Kowalczyk[1], A. Mirowska[5], M. Gryglas-Borysiewicz[6], J. Kossut[1],
M. Sawicki [1,3], A. Łusakowski [1], P. Bogusławski [1], T. Story[1, 4], K. Dybko [1, 4]

[1] *Institute of Physics, Polish Academy of Sciences, al. Lotników 32/46, 02668 Warsaw, Poland*

[2] *Center for Science and Innovation in Spintronics, Tohoku University, 2-1-1Katahira, Aoba-ku, Sendai 980-8577, Japan*

[3] *Research Institute of Electrical Communication, Tohoku University, 2-1-1Katahira Aoba-ku, Sendai 980-8577, Japan*

[4] *International Research Centre MagTop, Institute of Physics Polish Academy of Sciences, al. Lotników 32/46, 02668 Warsaw, Poland*

[5] *ENSEMBLE3 Centre of Excellence for Nanophotonics, Advanced Materials and Novel Crystal growth-based Technologies, Wólczyńska 133, 01919 Warsaw, Poland*

[6] *Faculty of Physics, University of Warsaw, Pasteura 5, 02093 Warsaw, Poland*

* Corresponding author: krolicka@ifpan.edu.pl

## CONTENTS

# S.I. Material characterization

## A) SIMS analysis

Pb$_{1-x-y}$Sn$_x$Cr$_y$Te samples are cut from the beginning and the end sections of the crystal ingots along the growth direction and subsequently subjected to SIMS analysis. The resulting concentrations of Sn and Cr are listed in Tab. S1 and are compared with the nominal composition. The examples of depth profiles (Fig. S1) showing elemental contribution as a function of distance from the edge indicate high homogeneity of the crystals.

Table S1. Results of SIMS analysis for Pb$_{1-x-y}$Sn$_x$Cr$_y$Te crystals with nominal composition $0.2 \leq x \leq 0.6$ and $y = 0.01$ investigated at the beginning and at the end of the crystals. Error of the composition determination is 0.005.

| nominal x | x begin. | x end | nominal y | y begin. | y end |
|---|---|---|---|---|---|
| 0.2 | 0.171 | 0.265 | 0.01 | 0.0017 | 0.0024 |
| 0.3 | 0.305 | 0.377 | 0.01 | 0.0012 | 0.0014 |
| 0.4 | 0.414 | 0.485 | 0.01 | 0.0078 | 0.0078 |
| 0.5 | 0.522 | 0.551 | 0.01 | 0.0077 | 0.0106 |
| 0.6 | 0.601 | 0.602 | 0.01 | 0.0063 | 0.0075 |

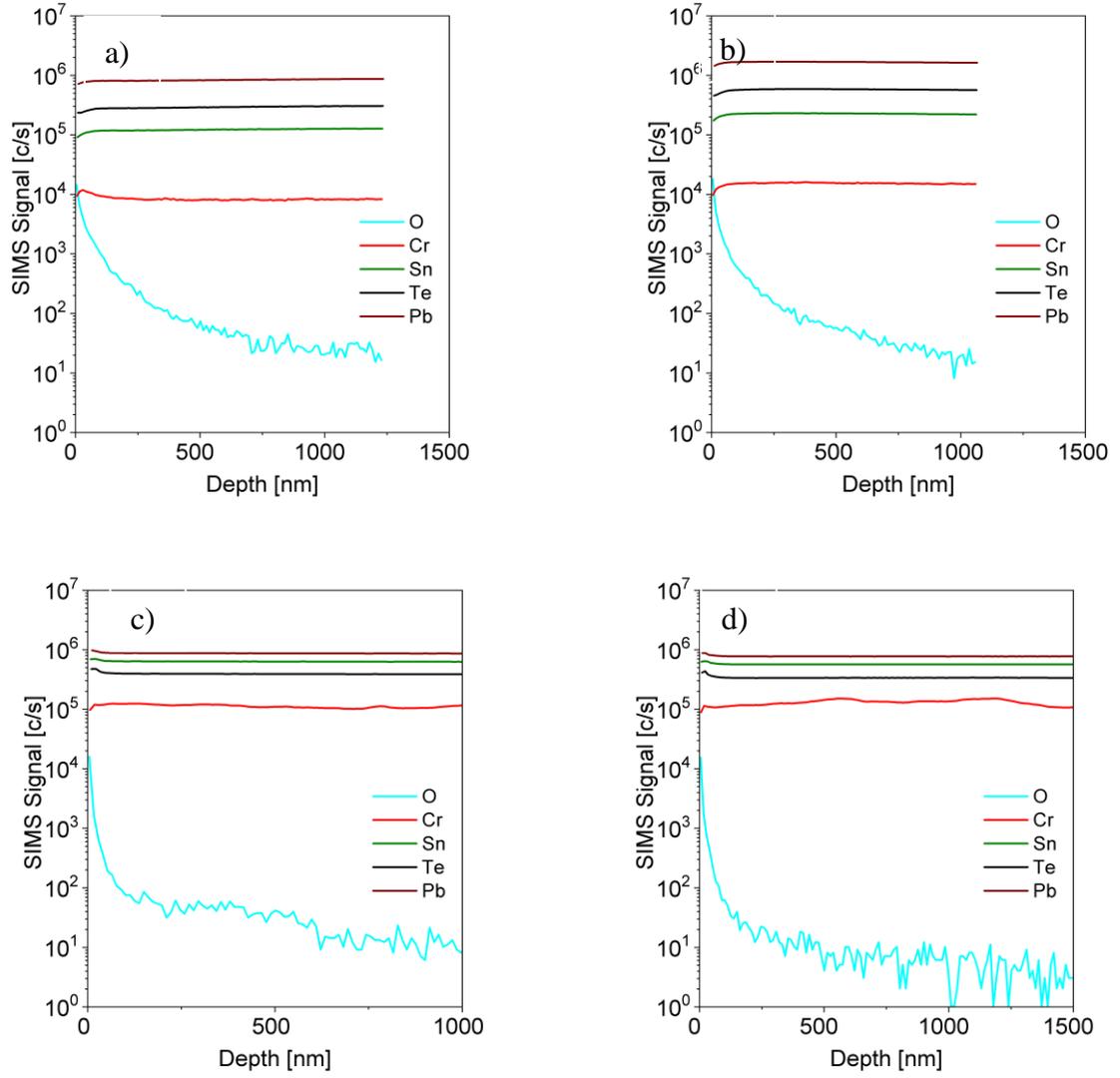

FIG. S1. SIMS depth profiles for (a) and (b) Pb$_{1-x-y}$Sn$_x$Cr$_y$Te $x = 0.2$, $y = 0.005$, (c) and (d) Pb$_{1-x-y}$Sn$_x$Cr$_y$Te $x = 0.6$, $y = 0.01$. Left panels show beginning section of the crystal ingots, right panels - the end sections of the crystal ingots.

## B) SEM/EDX analysis

The results of the EDX measurements show that $Pb_{1-x-y}Sn_xCr_yTe$ samples with Cr concentration lower than $y = 0.005$ are single-phase rock-salt materials. Above $y = 0.005$, the results reveal randomly distributed, elongated Cr-Te inclusions with length of about 10 - 50 μm and the widths of a few μm. Figure S2 shows SEM/EDX results of the sample with the nominal composition $Pb_{0.49}Sn_{0.5}Cr_{0.01}Te$. It follows from Fig. S2 (b-d), the inclusions contain only Cr and Te, while the signals from Pb and Sn are absent.

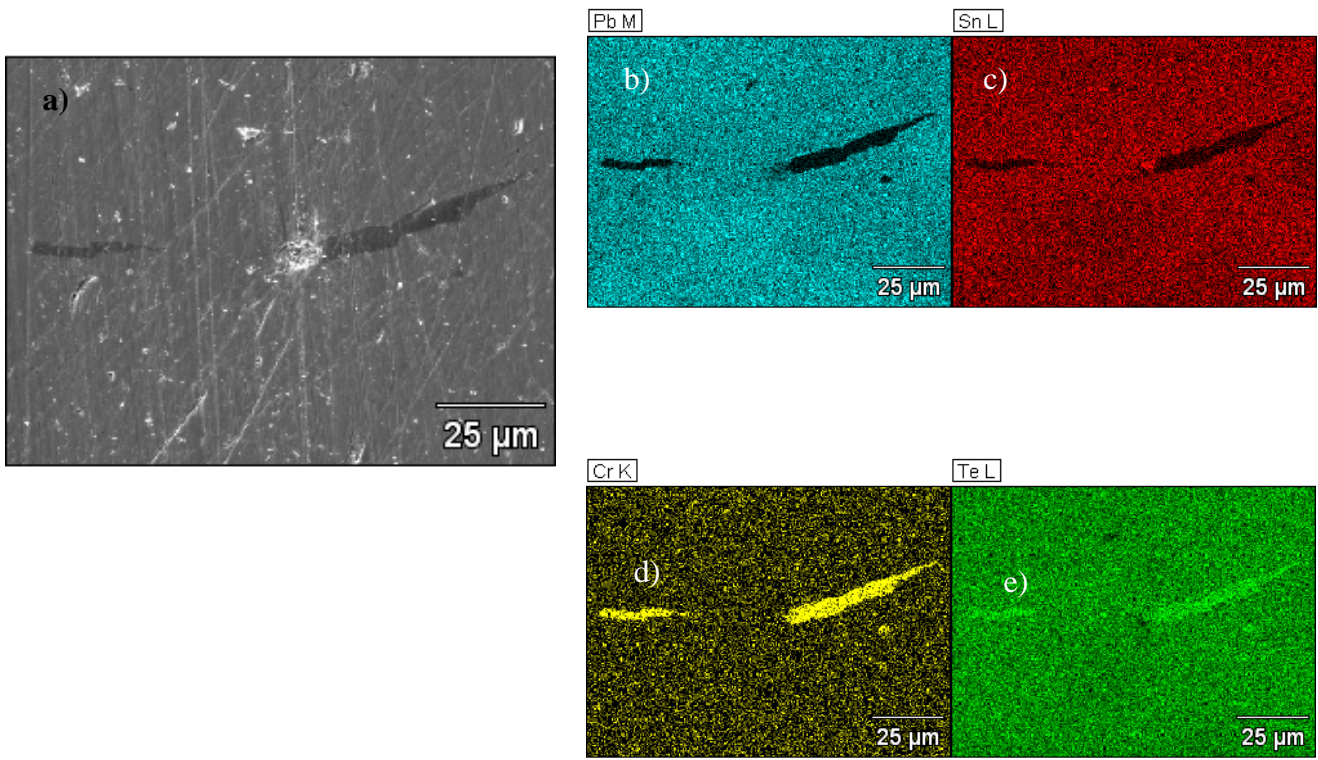

FIG. S2. Morphology of a $Pb_{0.49}Sn_{0.5}Cr_{0.01}Te$ specimen. (a) The sample surface (SEM image) and EDX maps revealing elongated μm-sized Cr-Te inclusions. Color intensities correspond to X-ray emission lines characteristic for the given elements: (b) $M$(Pb), (c) $L$(Sn), (d) $K$(Cr), (e) $L$(Te).

**C) XRD analysis**

The XRD studies allow to identify secondary crystallographic phases of Cr-Te precipitates. Figure S3 shows the XRD diffractograms of three samples cut from the beginning, the middle, and the end of an ingot, respectively. The Rietveld analysis reveals the presence of three types of Cr-Te inclusions, namely $Cr_2Te_3$ (Fig. S3 (a)), $Cr_5Te_6$ (Fig. S3 (b)) and $Cr_5Te_8$ (Fig. S3 (c)). The total contribution of the $Cr_2Te_3$, $Cr_5Te_6$ and $Cr_5Te_8$ inclusions in these samples is determined to be 0.6 wt.%, 0.3 wt.%, and 0.05 wt.%, respectively.

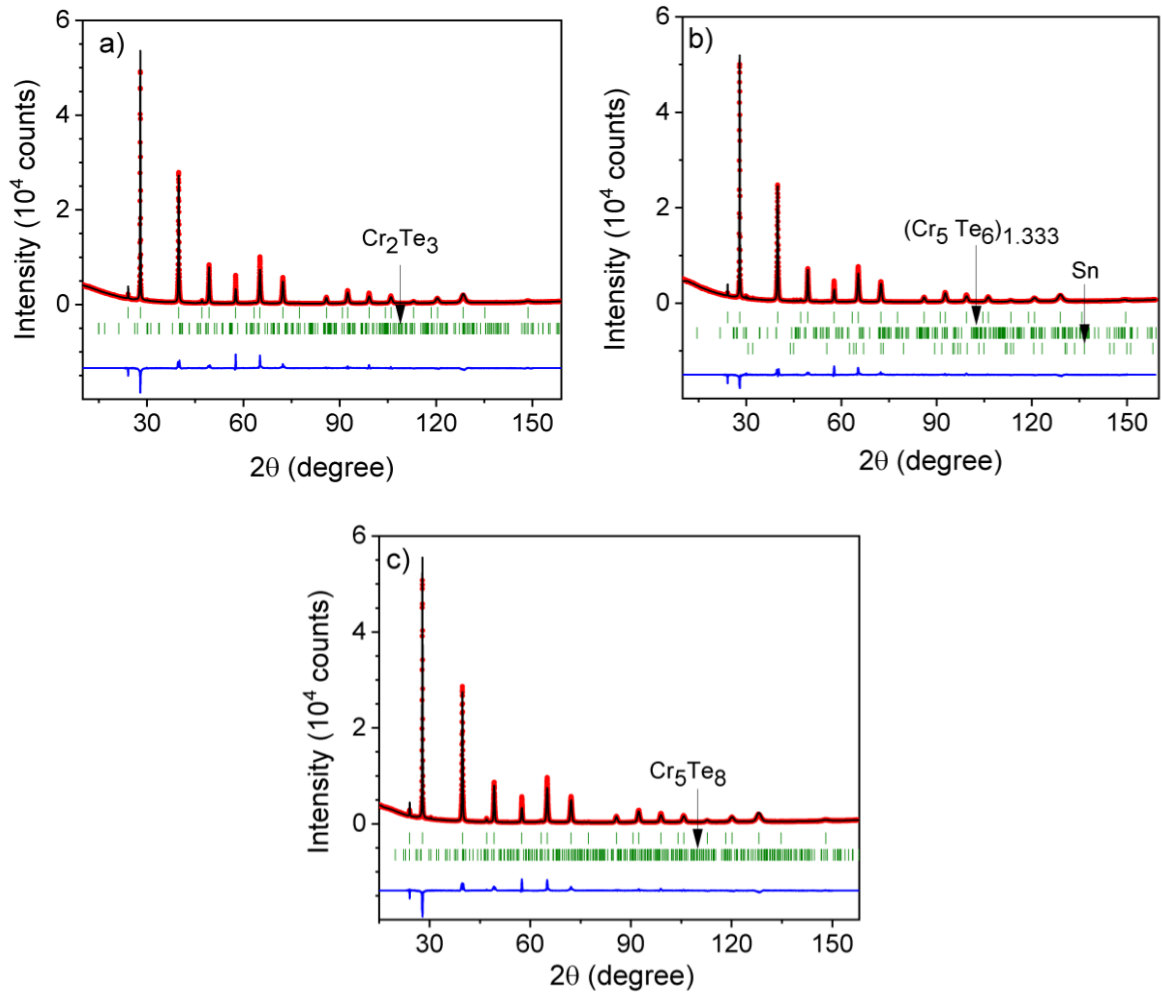

FIG. S3. Results of the X-Ray diffraction analysis of three samples cut from the beginning, the middle and the end sections of $Pb_{0.49}Sn_{0.5}Cr_{0.01}Te$ crystal ingot.

**D) Magnetometry**

SQUID magnetometry provides additional and more specific information regarding the Cr-Te inclusions detected by XRD. In Fig. S4 we show the remanent magnetization signal as a function of temperature and its derivative for a $Pb_{0.49}Sn_{0.5}Cr_{0.01}Te$ sample. From the minima of the temperature variation of the derivative of the remanent magnetization we can estimate critical temperatures of the secondary phases contained in the sample and, thus, deduce the ferromagnetic and/or antiferromagnetic character of the Cr-Te inclusions. The graph clearly shows three minima corresponding to three critical temperatures.

The literature data collected in Table S2 show a relatively large spread, therefore we only tentatively confirm contributions from four $Cr_nTe_m$ phases, namely $CrTe_3$, $Cr_2Te_3$, $Cr_5Te_6$ and $Cr_5Te_8$. In particular, the observed minimum at 65 K is close to $T_N = 55$ K of $CrTe_3$, the minimum at 170 K is close to $T_C$ of $Cr_2Te_3$ or to the critical temperatures of $Cr_5Te_8$, and the minimum at 265 K to $Cr_5Te_6$ or $Cr_5Te_8$. This tentative identification agrees with the XRD results (Fig. S3).

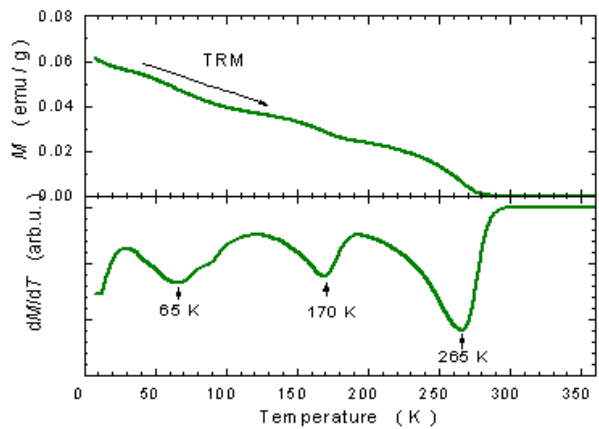

FIG. S4. Remanent magnetization, $M$ (upper panel) and its temperature, $T$, derivative (bottom panel) of $Pb_{0.49}Sn_{0.5}Cr_{0.01}Te$ sample.

Table S2. Various compounds based on the Cr-Te system with ferromagnetic ($T_C$) or anti-ferromagnetic ($T_N$) transition temperatures.

| Chromium telluride | $T_c$ [K] | $T_N$ [K] |
|---|---|---|
| $CrTe_3$ | - | 55 [1] |
| $Cr_2Te_3$ | 160-280 [2], 170-180 [3], 180 [4], 198 [5] | 130 [4] |
| $Cr_5Te_6$ | 320 [6] | 112 [6] |
| $Cr_5Te_8$ | 150 [7], 180-230 [8], 220 [9], 230 [10], 250 [11] | 160-180 [12], 180 [7] |

## S.II. $Pb_{1-x-y}Sn_xCr_yTe$ band structure parameters

The assumed band structure model relies on the non-parabolic two-band k·p approximation of energy versus *k*-vector dependence. We use temperature and composition variations of the experimental longitudinal ($m_l$) and transverse ($m_t$) effective masses for $Pb_{1-x}Sn_xTe$ from Ref. [13]. We assume the density of states effective mass in a single valley: $m^*=(m_l^* m_t^{*2})^{1/3}$. For each composition *x* we fit third degree polynomial. Exemplary expressions are presented in Table S3.

TABLE S3. Density of states effective mass for several tin compositions in $Pb_{1-x}Sn_xTe$ represented as third order polynomials: $m^*/m_0 = (a3 \cdot T^3 + a2 \cdot T^2 + a1 \cdot T + a0)$.

| x | a3[1/K³] | a2[1/K²] | a1[1/K] | a0 |
|---|---|---|---|---|
| 0 | -3.17×10⁻¹¹ | -1.17×10⁻⁸ | 7.79×10⁻⁵ | 0.052 |
| 0.056 | 2.79×10⁻¹⁰ | -9.95×10⁻⁸ | 8.37×10⁻⁵ | 0.045 |
| 0.08 | 8.83×10⁻¹⁰ | -3.83×10⁻⁷ | 1.26×10⁻⁴ | 0.041 |
| 0.12 | 9.08×10⁻¹¹ | -2.76×10⁻⁸ | 8.64×10⁻⁵ | 0.037 |
| 0.155 | 2.76×10⁻¹⁰ | -9.95×10⁻⁸ | 9.76×10⁻⁵ | 0.032 |
| 0.17 | -1.24×10⁻¹⁰ | 1.44×10⁻⁸ | 9.88×10⁻⁵ | 0.031 |

We also perform calculations to determine the band-gap dependencies on *T* and Sn content *x* basing on Varshni formula and the papers [13], [14]. The obtained relation describes satisfactorily the energy gap $E_{gap}$ for the whole range of temperatures and *x* values:

$$E_g(x,T)(meV) = 190 + 86x^2 - 576.3x + \frac{0.525 T^2}{T + 50}.$$

Temperature is in (K).

We assume that our $Pb_{1-x-y}Sn_xCr_yTe$ system is described by the following neutrality equation:

$$n_{Cr^{3+}} = n_{tot} - n_{Cr^{2+}}$$

where $n_{Cr^{3+}}$ is the concentration of ionized chromium donors, equivalent to the carrier concentration, $n_{Cr^{2+}}$ is the concentration of electrically neutral $Cr^{2+}$ ions and $n_{tot}$ is the total number of chromium dopant atoms. We neglect the presence of other defects and in particular the cation vacancies which act as acceptors. Since our analysis is limited solely to low Sn content ($x \leq 0.17$), the concentration of holes in the analyzed samples is about $10^{18}$ cm⁻³ (see Fig. 10 in the main text).

We use the formula for the concentration of ionized donors in the conduction band:

$$n_{Cr^{3+}} = \frac{N_v}{3\pi^2} \int_0^\infty -\left(\frac{\partial f_{F-D}(E)}{\partial E}\right) k^3(E) dE$$

where $N_V$ is the number of valleys in the conduction band, $f_{F-D}$ – the Fermi-Dirac distribution function, $E$ – energy in the band, $k$ – wave vector.

To obtain the concentration of neutral $Cr^{2+}$ ions we use the formula:

$$n_{Cr^{2+}} = \int f_{F-D} DOS_{Cr}(E) dE$$

where $DOS_{Cr}$ is the density of the resonant Cr states, represented by the Lorentz function and placed in the conduction band:

$$DOS_{Cr}(E) = n_{tot} \frac{W_{res}}{2\pi} \left[(E - E_{res})^2 + \left(\frac{W_{res}}{2}\right)^2\right]^{-1}$$

where $W_{res}$ is the width of the resonant chromium level, $E_{res}$ is the resonant energy calculated from the bottom of the conduction band.

Since $Pb_{1-x}Sn_xTe$ is characterized by relatively high dielectric constant, which effectively screens electrostatic interactions we have considered two other DOS function options namely, rectangular and Gaussian. These studies indicated that the best results are obtained when Lorentz function is used, surprisingly in accordance to systems where charge correlation effects are present with corresponding low dielectric constant [15,16].

The determined temperature dependence of $E_{res}$ is given by the formula: $E_{res} = E_{res}(0\ K) + (190 - E_{gap}(T))$ [meV]. In contrast to the previous models, which consider only a linear dependence of the Cr level on temperature, we find that a better description of the $E_{res}$ movement assumes that it follows the temperature dependent $E_{gap}$, including low temperature bowing of the band gap.

The conducted analysis allows to establish one consistent set of parameters for $Pb_{1-x-y}Sn_xCr_yTe$ in the whole composition range. According to the obtained results, the $Cr^{2+/3+}$ level in pure PbTe lies at the energy $E_{res}(0\ K) = 95$ meV, and its width $W_{res}$ is 3 meV.

The performed analysis allows also to determine the $Cr^{2+/3+}$ level location within the band structure of $Pb_{1-x}Sn_xTe$ as functions of temperature and composition. Figure S5 (a) shows the 3D diagram of the energy band structure of Cr doped $Pb_{1-x}Sn_xTe$ as a function of both $x$ and $T$ simultaneously. The composition for which the inversion of the bands occurs shifts from $x = 0.35$ at 4.2 K (Fig. S5 (c)) to $x = 0.75$ at 300 K. For higher Sn contents the band inversion point moves above room temperature.

Figure S5 (b - f) shows cross-sections of the band structure shown in panel (a) for several compositions of $Pb_{1-x-y}Sn_xCr_yTe$. With increasing $x$, the band gap gradually decreases until the zero-gap state is attained for $x = 0.35$ at $T = 4.2$ K, assuming concurrent band inversion model as stated in [13]. Fermi level becomes pinned in the vicinity of the band edge which is accompanied by a very low carrier concentration. When $x > 0.35$, the $L_6^+$ and $L_6^-$ states interchange their order and the band gap reopens again. Subsequently, the Cr impurity level enters the valence band and shifts downwards in the inverted band structure.

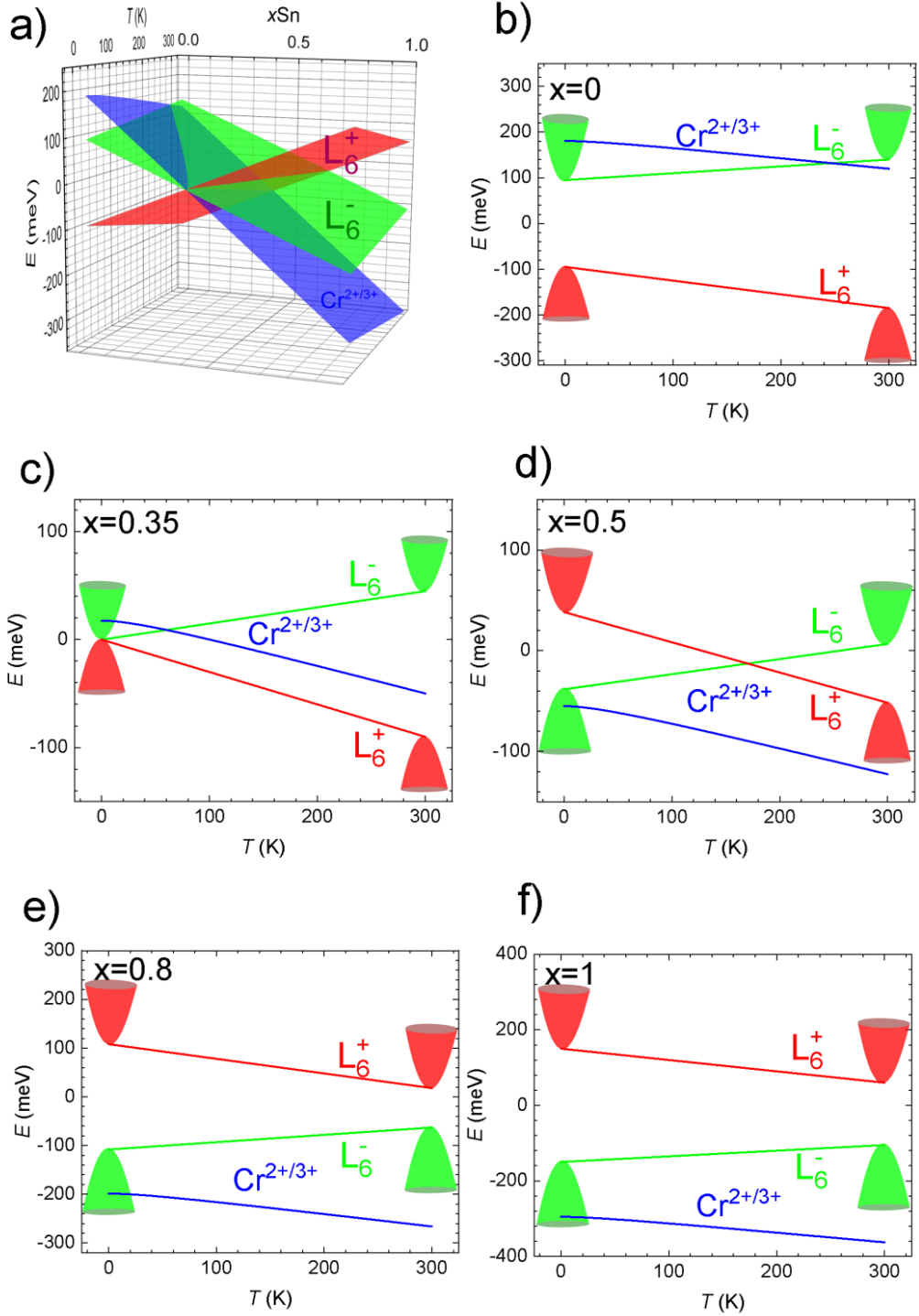

FIG. S5. Band structure of $Pb_{1-x-y}Sn_xCr_yTe$ compounds as a function of temperature and composition: (a) 3D diagram, (b - f) Cross-sections of (a) for several selected compositions. The $L_6^-$ and $L_6^+$ lines, marked with green and red through the whole temperature range, denote the conduction and valence bands symmetries. The blue solid line is the energy of $Cr^{2+/3+}$ resonant level. Conduction and valence bands are marked symbolically.

## S.III. Quantitative determination of Cr charge states from magnetization

Here we present decomposition of the total magnetization measured at the lowest temperature onto three paramagnetic components: $S = 3/2$ ($Cr^{3+}$), $S = 2$ ($Cr^{2+}$) and $S = 5/2$ ($Cr^{1+}$) (Fig. S6). We use the concentrations of particular chromium charge states determined in the main text (Fig. 11). Black dotted curve denotes the difference of the ferromagnetic contribution (ΔFM) measured at two lowest temperatures, specified in Fig. 11 of the main text. Since this difference should be close to zero, its position denotes the quality of determination of paramagnetic Cr components. The calculated paramagnetic contribution of $Cr^{3+}$ and $Cr^{2+}$ ions exceeds this difference, which justifies validity of the approach.

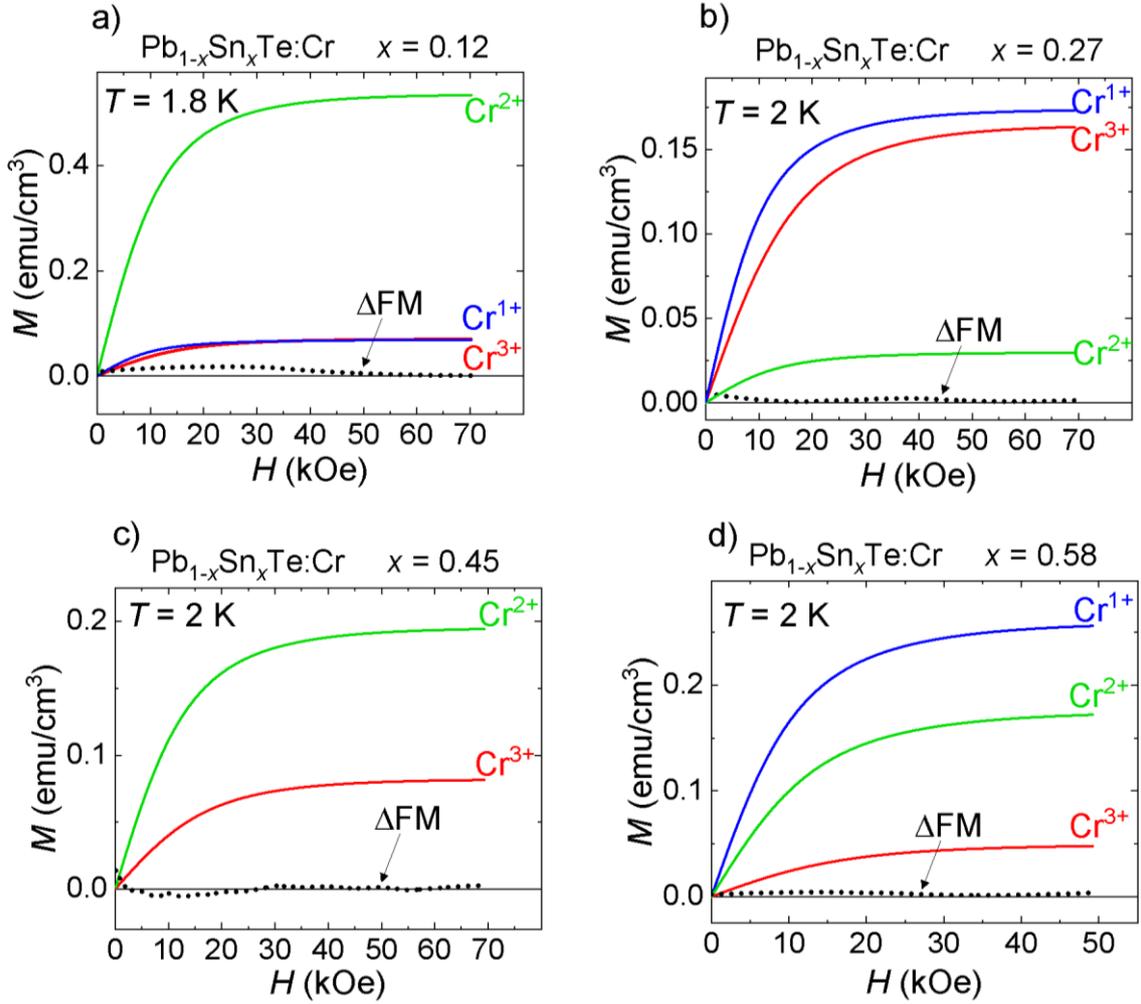

FIG. S6. Decomposition of the total magnetization, $M$, onto the paramagnetic components for several samples from Fig. 11 (main text). Paramagnetic components are marked as follows: $Cr^{3+}$ - red line, $Cr^{2+}$ - green line, $Cr^{1+}$ - blue line. Black dotted line denotes change of ferromagnetic contributions at two lowest temperatures $\Delta FM = M_{FM}(2K) - M_{FM}(5K)$, $M_{FM} = M_{TOT} - M_{PM}$, where $M_{TOT}$ – total measured signal and $M_{PM}$ – the paramagnetic contribution.